\documentclass[11pt,twoside]{article}
\usepackage{graphicx,floatflt,amssymb,pstricks}
\usepackage{times}
\usepackage{a4wide}
\usepackage{cite}

  \usepackage{fancyheadings}
  \advance \headheight by 3.0truept       

\newlength{\nseparation}
\newenvironment{nfigure}[1]
        {\begin{figure}[#1]\hrule\vspace{\nseparation}\par}
        {\vspace{\nseparation}\par \hrule \end{figure}}

\newcommand{\epm}[2]{
 \raisebox{-0.5ex}{\shortstack[l]{$\scriptstyle+#1$\\$\scriptstyle-#2$}}}

\newcommand{\bea}{\begin{eqnarray}}
\newcommand{\eea}{\end{eqnarray}}
\newcommand{\beq}{\begin{equation}}
\newcommand{\eeq}{\end{equation}}

\newcommand{\mm}{meson-antimeson}
\newcommand{\mmm}{\mm\ mixing}
\newcommand{\bb}{\ensuremath{B\!-\!\Bbar\,}}
\newcommand{\bbm}{\bb\ mixing}
\newcommand{\bbs}{\ensuremath{B_s\!-\!\Bbar{}_s\,}}
\newcommand{\bbms}{\bbs\ mixing}
\newcommand{\bbd}{\ensuremath{B_d\!-\!\Bbar{}_d\,}}
\newcommand{\bbmd}{\bbd\ mixing}
\newcommand{\bbq}{\ensuremath{B_q\!-\!\Bbar{}_q\,}}
\newcommand{\bbmq}{\bbq\ mixing}
\newcommand{\dd}{\ensuremath{D\!-\!\Dbar\,}}
\newcommand{\ddm}{\dd\ mixing}
\newcommand{\kk}{\ensuremath{K\!-\!\Kbar\,}}
\newcommand{\kkm}{\kk\ mixing}
\newcommand{\dg}{\ensuremath{\Delta \Gamma}}
\newcommand{\dm}{\ensuremath{\Delta M}}

\newcommand{\eq}[1]{Eq.~(\ref{#1})}
\newcommand{\eqsand}[2]{Eqs.~(\ref{#1}) and (\ref{#2})}
\newcommand{\eqsto}[2]{Eqs.~(\ref{#1}--\ref{#2})}

\newcommand{\Dbar}{\,\overline{\!D}}
\newcommand{\Bbar}{\,\overline{\!B}}
\newcommand{\Kbar}{\,\overline{\!K}}
\newcommand{\Mbar}{\,\overline{\!M}}
\newcommand{\no}{\nonumber}
\newcommand{\nn}{\nonumber \\}
\newcommand{\ov}[1]{\overline{#1}}
\newcommand{\lt}{\left}
\newcommand{\rt}{\right}

\newcommand{\mev}{\mbox{MeV}}
\newcommand{\gev}{\mbox{GeV}}

\newcommand{\fig}[1]{Fig.~\ref{#1}}
\newcommand{\ds}{\displaystyle}
\newcommand{\imag}{\mathrm{Im}\,}
\newcommand{\real}{\mathrm{Re}\,}

\newcommand{\lqcd}{\Lambda_{\rm QCD}}
\newcommand{\bra}[1]{\ensuremath{\langle #1 |}}
\newcommand{\ket}[1]{\ensuremath{| #1 \rangle }}

\newcommand{\e}{\epsilon}

\newcommand{\gtf}{\ensuremath{\Gamma (M(t) \rightarrow f )}}
\newcommand{\gbtf}{\ensuremath{\Gamma (\Mbar{}(t) \rightarrow f )}}
\newcommand{\gtfb}{\ensuremath{\Gamma (M(t) \rightarrow \ov{f} )}}
\newcommand{\gbtfb}{\ensuremath{\Gamma (\Mbar{}(t) \rightarrow \ov{f} )}}
\newcommand{\gtfcp}{\ensuremath{\Gamma (M(t) \rightarrow f_{\rm CP} )}}
\newcommand{\gbtfcp}{\ensuremath{\Gamma (\Mbar{}(t) \rightarrow f_{\rm CP} )}}

\newcommand{\guntf}{\ensuremath{\Gamma  [f,t] }}
\newcommand{\guntfb}{\ensuremath{\Gamma  [\ov{f},t] }}

\newcommand{\DorDbar}{\raisebox{7.7pt}{$\scriptscriptstyle(\hspace*{9.5pt})$}
  \hspace*{-11.7pt}\!\Dbar}

\begin{document}
\thispagestyle{plain}
\parbox[t]{0.4\textwidth}{TTP09-07}
\hfill March 2009 \\

\begin{center}
\boldmath
{\huge \bf Three Lectures on\\[4mm] 
            Meson Mixing and CKM phenomenology\footnote{Contribution to 
the \emph{Helmholtz International Summer School
``Heavy quark physics''}, 
Bogoliubov Laboratory of Theoretical Physics,
Dubna, Russia, August 11-21, 2008.}}\\
\unboldmath
\vspace*{1cm}
\renewcommand{\thefootnote}{\fnsymbol{footnote}}
Ulrich Nierste \\
\vspace{10pt}
{\small
              {\em Institut f\"ur Theoretische Teilchenphysik\\
               Universit\"at Karlsruhe\\ 
               Karlsruhe Institute of Technology, 
               \\ D-76128 Karlsruhe, Germany}} \\
\normalsize
\end{center}

\begin{abstract}
I give an introduction to the theory of meson-antimeson mixing, aiming at 
students who plan to work at a flavour physics experiment or intend to
do associated theoretical studies. I derive the formulae for the time
evolution of a neutral meson system and show how the mass and width
differences among the neutral meson eigenstates and the CP phase in mixing
are calculated in the
Standard Model. Special emphasis is laid on CP violation, which is
covered in detail for \kkm, \bbmd\ and \bbms. I explain the constraints 
on the apex $(\ov \rho,\ov \eta)$ of the unitarity triangle implied by
$\e_K$, $\dm_{B_d}$, $\dm_{B_d}/\dm_{B_s}$ and various mixing-induced 
CP asymmetries such as $a_{\rm CP} (\ov B_d \to J/\psi K_{\rm
  short})(t)$. 
The impact of a future measurement of CP violation in flavour-specific   
$B_d$ decays is also shown.  
\end{abstract}

\tableofcontents

\section{First lecture: A big-brush picture}
\subsection{Mesons, quarks and box diagrams}\label{sec:mqb}
The neutral $K$, $D$, $B_d$ and $B_s$ mesons are the only hadrons which
mix with their antiparticles.  These meson states are flavour
eigenstates and the corresponding antimesons $\Kbar$, $\Dbar$,
$\Bbar_d$ and $\Bbar_s$ have opposite flavour quantum numbers:
\bea%
&& K \sim \ov s d, \qquad D \sim c \ov u, \qquad B_d \sim \ov b d,
\qquad B_s \sim \ov b s, \nn && \Kbar \sim s \ov d, \qquad \Dbar \sim
\ov c u, \qquad \Bbar_d \sim b \ov d, \qquad \Bbar_s \sim b \ov s,
\label{fqn} \eea%
Here for example ``$ B_s \sim \ov b s $'' means that the $ B_s$ meson
has the same flavour quantum numbers as the quark pair $( \ov b ,s)$,
i.e.\ the beauty and strangeness quantum numbers are $B=1$ and $S=-1$,
respectively. The meson states in \eq{fqn} are also eigenstates of the
strong and electromagnetic interactions. As long as we neglect the weak
interaction, they are also mass eigenstates, with the same mass for
meson and antimeson.  In the Standard Model (SM) all interaction
vertices conserve flavour, except for the couplings of W bosons to
fermions.\footnote{Strictly speaking, this statement assumes that the
  so-called unitary gauge for the weak gauge bosons is adopted.  The
  unphysical charged pseudo-Goldstone bosons, which appear in other
  gauges, also have flavour-changing vertices. Changing the gauge
  shuffles terms between the pseudo-Goldstone bosons and the
  longitudinal components of the gauge bosons.} The piece of the SM
Lagrangian which describes the W couplings to quarks reads \beq {\cal
  L}_W = \frac{g_w}{\sqrt{2}} \sum_{j,k=1,2,3} \lt[ V_{jk} \,
\ov{u}_{jL} \, \gamma^{\mu} d_{kL} \, W^{+}_{\mu} + V_{jk}^* \,
\ov{d}_{kL} \, \gamma^{\mu} u_{jL} \, W^{-}_{\mu} \rt] .
 \label{wex}
\eeq
Here $g_w$ is the weak coupling constant and $V$ is the $3\times 3$ unitary 
\emph{Cabibbo-Kobayashi-Maskawa (CKM) matrix}: 
\bea%
V &=& \left( \begin{array}{ccc}
                V_{ud} & V_{us} & V_{ub} \\
                V_{cd} & V_{cs} & V_{cb} \\
                V_{td} & V_{ts} & V_{tb}
        \end{array} \right). \label{defv}   
\eea%
In \eq{wex} I have further used the notations $(d_1,d_2,d_3)=(d,s,b)$ and
$(u_1,u_2,u_3)=(u,c,t)$. The W boson only couples to the left-handed
components of the quark fields as indicated by the subscript ``L'' in
\eq{wex}.  At fourth order in the weak coupling we can change the flavour
quantum numbers by two units and obtain transitions between mesons and
antimesons. The corresponding Feynman diagrams are shown in \fig{fig:boxes}.
\begin{nfigure}{tbp}  
\begin{center}
\begin{tabular}{c@{\hspace{1cm}}c}
\includegraphics[width=0.4\textwidth]{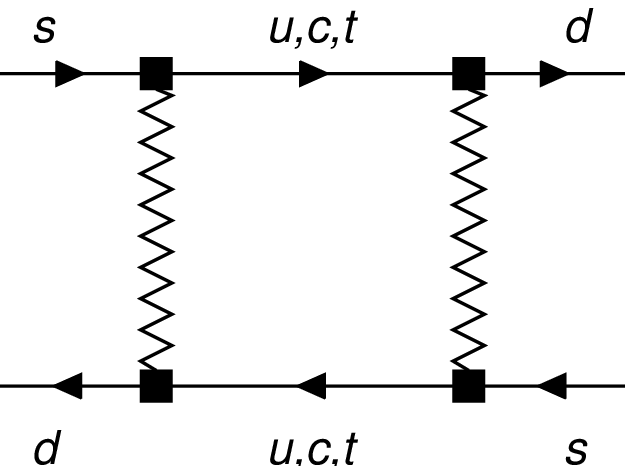} &
\includegraphics[width=0.4\textwidth]{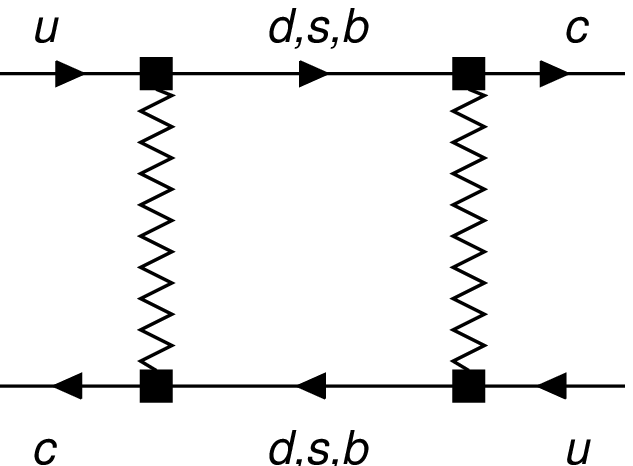} \\[3mm]
\includegraphics[width=0.4\textwidth]{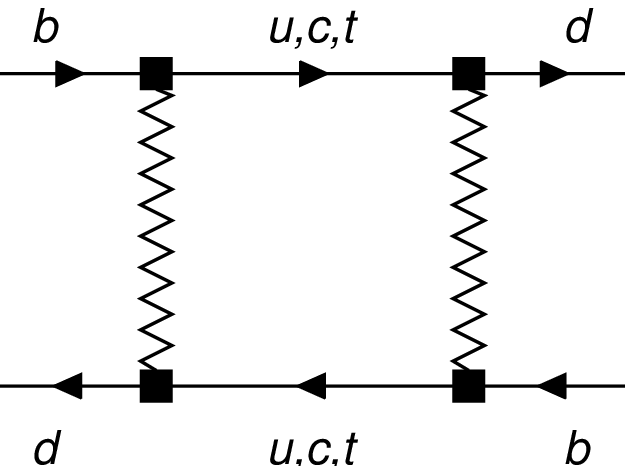} &
\includegraphics[width=0.4\textwidth]{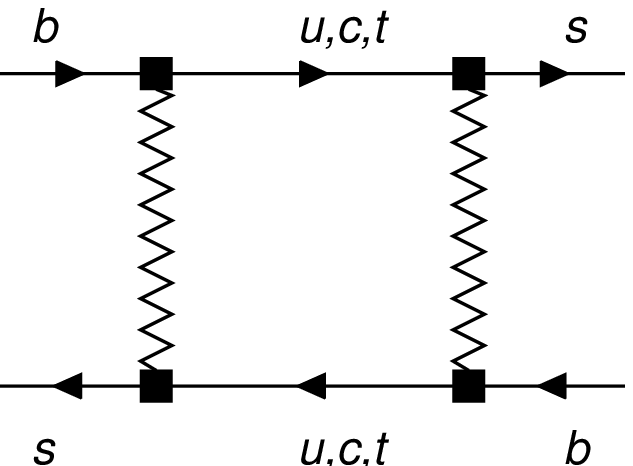} 
\end{tabular}
\end{center}
\caption{Box diagrams for \kk, \dd, \bbd\ and \bbms. The zigzag lines
  represent W bosons. For each process there is also a second box diagram, 
  obtained by a 90$^\circ$ rotation. }\label{fig:boxes} 
\end{nfigure}
These $|\Delta F|=2$ diagrams, where $F$ denotes the appropriate flavour
quantum number $F=S$, $C$ or $B$, represent the lowest non-vanishing
contribution to the transition matrix element $\Sigma_{12}$ defined by
\bea%
-i (2\pi)^4 \delta^{(4)}(p_M- p_{\ov{\!M}}) \Sigma_{12} &=&
\frac{\bra{M(\vec p_M)} S \ket{\,\ov{\!M} (\vec p_{\ov{\!M}})} }{2 M_M}
\label{defa12} \eea%
with the S-matrix $S$ and the generic notation $M=K$,$D$,$B_d$ or $B_s$.
(The notation $\Sigma_{12}$ refers to the quantum-mechanical two-state
system with $\ket 1=\ket M$ and $\ket 2=\ket{\ov M}$.) I comply with the
standard relativistic normalisation of the meson states, $\bra{M(\vec
  p\,{}^\prime)} M(\vec p )\rangle = 2E \, (2\pi)^3 \delta^{(3)} (\vec
p\,{}^\prime-\vec p )$. The meson mass $M_M=\sqrt{E^2-\vec p\,{}^2}$ in
the denominator in \eq{defa12} is introduced for later convenience.  In
terms of the Hamiltonian (density) $ H_{\rm int}^{SM}(x)=-{\cal L}_{\rm
  int}^{SM}(x) $, which encodes all interactions of the SM, the S-matrix
is given by the usual time-ordered exponential%
\bea%
S &=& \mathbf{T} e^{-i \int d^4 x H_{\rm int}^{\rm SM}(x)}
 . \label{defs} 
\eea%
In order to link \eqsand{defa12}{defs} to the diagrams of
\fig{fig:boxes} we must consider the contribution from ${\cal L}_W$ in
\eq{wex} to $ - H_{\rm int}^{SM} $ and expand the time-ordered
exponential in \eq{defs} to order $g_w^4$. The determination of this
term amounts to the calculation of the two contributing box diagrams
with the usual Feynman rules of the weak interaction. To this point we
have only used standard text-book quantum field theory, noting an
important omission: No effect of the strong interaction has been taken
into account by now. Most importantly, we do not know yet how to take
care of quark confinement, which forces the external quarks in the
diagrams of \fig{fig:boxes} to form mesons. As an important feature,
Quantum Chromodynamics (QCD) behaves very differently at short and long
distances: At short distances (probed by large energies) the QCD
coupling constant $g_s$ is small and we can apply perturbation theory
\cite{gpw}, just as we did with the weak interaction. That is, effects
of short-distance QCD can be included by adding gluons to the diagrams
in \fig{fig:boxes}. At large distances, corresponding to low energies,
QCD is non-perturbative and one must resort to different methods, such
as lattice gauge theory or QCD sum rules.  Long-distance QCD is also
referred to as \emph{hadronic physics}, because its degrees of freedom
are hadrons rather than quarks and gluons.  In many cases the associated
theoretical uncertainties are the main obstacle in the relation between
measured quantities and the fundamental parameters of nature encoded in
the Lagrangian ${\cal L}$. Theorists pursue a two-fold strategy to deal
with hadronic uncertainties: On one hand they try to refine
non-perturbative methods such as lattice gauge theory. On the other hand
they try to identify quantities in which hadronic uncertainties are
small or even absent or look for ways to eliminate hadronic uncertainties
through clever combinations of different observables. We will encounter
both strategies in our discussion of meson-antimeson mixing.  Weak
processes of hadrons involve several largely-separated energy
scales.\footnote{I use natural (or Planck) units with $\hbar=c=1$, so
  that masses and momenta have units of \gev.} For example, in \bbm\ we
encounter $m_t > M_W \gg m_b \gg \lqcd$, where $ \lqcd\sim 0.4\,\gev$ is
the fundamental scale of the strong interaction governing e.g.\ the size
of binding energies.  In order to correctly calculate $\Sigma_{12}$ we
must separate the different scales from each other and apply different
computational methods to large and small energy scales. However, without 
detailed understanding of the strong interaction we can roughly assess the
relative importance of the contributions from the different
internal quark flavours in \fig{fig:boxes}: In the case of \bbmd\ and
\bbms\ one finds that the box diagram with internal top quarks vastly
dominates over the diagrams with light quarks, because the result of the
diagram grows with the internal quark mass. For \kkm\ and \ddm\ no
such estimate is possible, because the contribution with the heaviest 
quark is suppressed by small CKM elements. 
 
Owing to $\Sigma_{12}\neq 0$, $M$ and $\,\ov{\!M}$ mix and are no more
mass eigenstates. The latter are obtained by diagonalising the $2\times
2$ matrix $\Sigma_{ij}$, where 
\begin{eqnarray}
  -i (2\pi)^4 \delta^{(4)}(p_i^\prime- p_j)  \Sigma_{ij} 
     &=& \frac{\bra{i, \vec p_i{}^\prime} 
               S^{\rm SM}
               \ket{j, \vec p_j} 
              }{2 M_M} \label{defaij} 
\end{eqnarray}
with $\ket{1, \vec p_1}= \ket{ M (\vec p_1)}$ and $\ket{2, \vec p_2}=
\ket{\,\ov{\!M} (\vec p_2)}$ generalises \eq{defa12}.  We list two
important aspects of meson-antimeson mixing:
\begin{itemize}
\item[i)] The two mass eigenstates are linear
          combinations of $M$ and $\,\ov{\!M}$. The degeneracy is lifted and 
          we can denote the two mass eigenstates by $M_H$ and $M_L$, where 
          ``$H$'' and ``$L$'' stand for ``heavy'' and ``light'',
          respectively. $M_H$ and $M_L$ not only differ in their masses, 
          but also in their lifetimes.   
\item[ii)] If we produce a meson $M$ at some time $t=0$, 
           the corresponding state will 
           evolve into a superposition of $M$ and $\,\ov{\!M}$ at later times
           $t>0$. One observes  meson-antimeson oscillations.  
\end{itemize}    
We will calculate the differences among the masses and decay widths in
the second and third lectures.  Studies of neutral Kaons mainly exploit
property i), while the mixings of the other three neutral meson systems
are investigated through property ii). The reason for the Kaon's special
role here is the vast lifetime difference between $K_H$ and $K_L$. The
former state, usually denoted as $K_{\rm long}$, lives roughly 500 times
longer than $K_L=K_{\rm short}$, so that one can easily produce a
$K_{\rm long}$ beam. For $D$, $B_d$ and $B_s$ mesons the width
differences are much smaller than the average decay width of the two
eigenstates and this method is not feasible. The identification of the
meson (discriminating between $M$ and $\,\ov{\!M}$) needed to track the
meson-antimeson oscillations is called \emph{flavour tagging}.  To
observe the oscillations the mesons must move sufficiently fast in the
detector.  Modern B factories, which produce $(B_d,\Bbar_d{})$ pairs via
the $\Upsilon(4S)$ resonance, have therefore asymmetric beam energies,
so that the centre-of-mass frame (coinciding with the rest frame of the
$\Upsilon(4S)$) moves with respect to the laboratory frame.  At
hadron colliders studies of meson-antimeson oscillations profit from the
large boost of the produced mesons.  Tevatron and LHC are especially
powerful for $B_s$ physics, because the \bbs\ oscillations are very
rapid.

\subsection{A bit of history}
Meson-antimeson mixings belong to the class of \emph{flavour-changing
  neutral current (FCNC)}\ processes, which involve different flavours
with the same electric charge. Since in the SM such processes are
forbidden at tree-level, they are sensitive to new heavy particles
appearing as virtual particles in loop diagrams. Historically, the first
new particle predicted from the consideration of FCNCs was the charm
quark, which was needed to eliminate large tree-level FCNC couplings in
conflict with experiment \cite{gim}. Subsequently, the rough size of the
charm quark mass $m_c$ was predicted from the size of the mass
difference $\dm_K=M_H-M_L$ in the neutral Kaon system \cite{gl}.  A
great success story of flavour physics has been the exploration of the
discrete symmetries charge conjugation ($C$), parity ($P$) and time
reversal ($T$).  Charged Kaon decays had revealed in 1956 that $P$ and
$C$ are not conserved by the weak interaction, while physicists kept
their faith in a good $C\!P$ symmetry. If $C\!P$ were conserved, we could
assign $C\!P$ quantum numbers to $K_{\rm long}$ and $K_{\rm short}$. The
latter meson was observed to decay into a two-pion state, and each pion
is $C\!P$-odd and contributes a factor of $-1$ to the total $C\!P$ quantum
number (which is multiplicative).  A further contribution to the $C\!P$
quantum number of a two-particle state stems from the angular momentum:
States with orbital angular momentum quantum number $l$ involve the
spherical harmonic $Y^l_m(\vec n)$, where $\vec n=\vec p/|\vec p|$ and 
$\vec p$ is the relative momentum of the two particles considered.  
Since $Y^l_m(\vec n)=(-1)^l
Y^l_m(- \vec n)$, states with odd $l$ have $P$ and $C\!P$ quantum numbers
$-1$, while those with even $l$ are even under $P$ and $C\!P$. Since the
decaying Kaon has no spin and the total angular momentum is conserved in
any decay process, the two pions in the final state have have $l=0$ in
the Kaon rest frame.  (In general the spin wave function also matters,
but pions have spin zero.)  In total we find that the two-pion state is
$C\!P$-even.  Now $K_{\rm long}$ was only seen to decay into three pions, so
that this meson was believed to be $C\!P$-odd.  In fact, its long lifetime
stems from the kinematical suppression of the decay into the $C\!P$-odd
three--pion state. To understand that a three-pion state is always
$C\!P$-odd, first note that we get a contribution of $(-1)^3=-1$ from the
intrinsic $C\!P$ quantum numbers of the three pions. Next pick any two of
the pions and call there relative orbital angular momentum quantum
number $l_1$.  Likewise we denote the quantum number for the relative
orbital angular momentum between this pair and the third pion by $l_2$.
One of the selection rules for the addition of angular momenta implies
that the total quantum number $l$ satisfies $l \geq |l_1-l_2|$. Since
$l=0$, this means that $l_1=l_2$ and the ``orbital'' contribution to the
$C\!P$ quantum number is $(-1)^{l_1+l_2}=(-1)^{2 l_1}=1$. Thus the
three-pion state is $C\!P$-odd, irrespective of the value of $l_1$.
   
In 1964 the decay $K_{\rm long}\to \pi\pi$ was observed, establishing $C\!P$ violation
\cite{ccft}. The two-generation Standard Model, whose construction was
completed later in that decade \cite{gsw}, could not accommodate this
phenomenon: We will see below that $C\!P$-violating interactions of quarks
necessarily involve complex couplings.  While the $V_{jk}$'s in \eq{wex} are a
priori complex, one can render them real in the two-generation SM by
transforming the quark fields as
\bea%
  d_j \to e^{i \phi^d_j} d_j,&& \qquad\qquad 
  u_k \to e^{i \phi^u_k} u_k . \label{fph}
\eea%
with appropriate phases $\phi^d_j$ and $\phi^u_k$.  The net effects of these
rephasings are the replacements of the $V_{jk}$'s by
\bea%
   \qquad && V_{jk}  e^{i (\phi^d_j-\phi^u_k)} .\label{vph}
\eea%
These expressions involves three independent phases and we may choose e.g.\ 
$\phi^d_1-\phi^u_1$, $\phi^d_1-\phi^u_2$ and $\phi^d_2-\phi^u_1$ in such a way
that the three complex phases of a unitarity $2\times2$ matrix are eliminated,
arriving at the real Cabibbo matrix.  In 1973 Kobayashi and Maskawa have
pointed out that a physical $C\!P$-violating phase persists in the quark mixing
matrix, if there are at least three generations \cite{km}: A unitary $3\times
3$ matrix has 6 complex phases while we have only 5 phase differences
$\phi^d_j-\phi^u_k$ at our disposal. The finding of Kobayashi and Maskawa was
largely ignored at that time and only appreciated after the third fermion
generation was experimentally established. In 1987 the ARGUS experiment at
DESY observed \bbmd, at an unexpectedly large rate \cite{argus}. This
finding was the first hint at a truly heavy top quark, which enters the lower
left box diagram of \fig{fig:boxes}. 

\boldmath
\subsection{$C\!P$ violation}
\unboldmath
The last stroke of the brush is devoted to $C\!P$ violation. Defining 
\begin{eqnarray}
CP  \ket{\, M (\vec p_{\ov{\!M}})} \;=\; 
    - \ket{\;\ov{\!M} (-\vec p_{\ov{\!M}})},\qquad\qquad && 
CP  \ket{\;\ov{\!M} (\vec p_{\ov{\!M}})} \;=\; 
    - \ket{\, M (-\vec p_{\ov{\!M}})} \label{defcpm}
\end{eqnarray}
we first look at decays $M\to f_{\rm CP}$ and  $\ov{\!M} \to f_{\rm CP}$,
where $f_{\rm CP}$ is a $C\!P$ eigenstate:
\begin{eqnarray}
  CP \ket{f_{\rm CP}} &=& \eta_{\rm CP} \ket{f_{\rm CP}} \label{defcpeig}
\end{eqnarray}
with $\eta_{\rm CP}=\pm 1$.  The $C\!P$ operator appearing in 
\eqsand{defcpm}{defcpeig} is unitary, i.e.\ $(C\!P)^{-1}=(C\!P)^\dagger$. 
To get an idea of the importance of \mmm\
for the study of $C\!P$ violation we first assume that $M$ and
$\ov{\!M}$ do not mix.  We could still measure the decay rates of the
$C\!P$-conjugate processes $M\to f_{\rm CP}$ and $\ov{\!M} \to f_{\rm
  CP}$.  If we find them different we establish \emph{direct $C\!P$
  violation}\ (often called \emph{$C\!P$ violation in decay}).  However,
it is very difficult to relate a direct $C\!P$ asymmetry to a
fundamental $C\!P$ phase in ${\cal L}$: A non-zero direct $C\!P$
asymmetry also requires final state interaction related to the
rescattering process $M\to f^\prime_{\rm CP} \to f_{\rm CP}$.
Rescattering leads to $C\!P$-conserving complex phases in the decay
amplitude. In the absence of such phases the amplitudes of $M\to f_{\rm
  CP}$ and $\ov{\!M} \to f_{\rm CP}$ would simply be related by complex
conjugation since all phases would switch sign under $C\!P$. But then the
two decay amplitudes would have the same magnitude leading to identical
decay rates.  For $M=D,B_d,B_s$ this hadronic rescattering process is
mainly inelastic and intractable with present theoretical methods.

But thanks to \mmm\ we can study meson states which are superpositions of
$\ket{M}$ and $\ket{\ov{\!M}}$. The mass eigenstates $\ket{M_H}$ and
$\ket{M_L}$ are linear combinations of $\ket{M}$ and $\ket{\ov{\!M}}$:
\bea%
\ket{M_L} &=&
    p \ket{M} + q \ket{\, \ov{\!M} } \,,  \nn
\ket{M_H} &=&
    p \ket{M} - q \ket{\, \ov{\!M} } \,,
\label{defpq}
\eea%
with $\lt|p\rt|^2+\lt|q\rt|^2 = 1$. We can calculate $p$ and $q$ from
the box diagrams in \fig{fig:boxes} and will do so in the following
sections. A commonly used shorthand notation for decay amplitudes is
\bea%
A_f = A(M\to f) = \bra{f} S \ket{M} 
,&&\qquad\qquad \ov A_f = A(\,\ov{\!M} \to f ) = 
               \bra{f} S \ket{\,\ov{\!M}}  .
\label{defaf}%
\eea%
A key quantity to study $C\!P$ violation is the combination%
\bea%
\lambda_f &=& \frac{q}{p}\, \frac{\ov{A}_f}{A_f} . \label{deflaf}%
\eea%
$\lambda_f$ encodes the essential feature of the interference of the $M\to f $
and $\,\ov{\! M} \to f$ decays, the relative phase between $q/p$ (from \mmm )
and $\ov{A}_f/A_f$ (stemming from the studied decay).  In a first application,
I discuss the decays of neutral Kaons into two charged or neutral pions.  A
neutral $K$ or $\ov K$ meson state is a superposition of $K_H=K_{\rm long}$
and $K_L=K_{\rm short}$. At short times the decays of the $K_{\rm short}$
component of our Kaon beam will vastly dominate over the $K_{\rm long}$ decays
and one can access the decay rates $\Gamma (K_{\rm short} \to \pi \pi)$ for
$\pi\pi=\pi^+\pi^-,\pi^0\pi^0 $. At large times, say, after 10 times the
$K_{\rm short}$ lifetime, our beam is practically a pure $K_{\rm long}$ beam
and we can study the $C\!P$-violating $\Gamma (K_{\rm long} \to \pi \pi)$
decays.  It is advantageous to switch to the eigenbasis of strong
isospin $I$:%
\bea%
\ket{\pi^0 \pi^0 } & = & \sqrt{\frac{1}{3}} \, \ket{\lt(\pi \pi
  \rt)_{I=0}} - \sqrt{\frac{2}{3}} \, \ket{\lt(\pi \pi \rt)_{I=2}} \,,
\nn \ket{\pi^+ \pi^- } & = & \sqrt{\frac{2}{3}} \, \ket{\lt(\pi \pi
  \rt)_{I=0}} + \sqrt{\frac{1}{3}} \, \ket{\lt(\pi \pi \rt)_{I=2}} \,,
\no \eea%
The strong interaction respects strong-isospin symmetry to an accuracy
of typically 2\%, so that we can neglect any rescattering between the
$I=0$ and $I=2$ states. Consequently, no \emph{direct}\ $C\!P$ violation
contributes to the famous $C\!P$-violating quantity%
\beq%
\e_K \equiv \frac{ \bra{(\pi\pi)_{I=0}} K_{\rm long} \rangle }{
  \bra{(\pi\pi)_{I=0}} K_{\rm short} \rangle }
\label{ek} .
\eeq%
Abbreviating $A_0\equiv A_{(\pi\pi)_{I=0}}$, $\ov A_0\equiv \ov
A_{(\pi\pi)_{I=0}}$ and (see \eq{deflaf}) $\lambda_0 \equiv
\lambda_{(\pi\pi)_{I=0}}$ we insert \eq{defpq} into \eq{ek} and readily find
\beq%
\e_K = \frac{1-\lambda_0}{1+\lambda_0} \,.
\label{ekla}
\eeq%
The experimental value \cite{pdg}
\beq%
\e_K^{\rm exp} = e^{i\, \phi_{\e}} \, (2.23 \pm 0.01) \times
    10^{-3} \qquad \qquad \mbox{with } \quad
  \phi_{\e} \; =\;  (0.967 \pm 0.001 )\, \frac{\pi}{4} 
\,. \label{ekexp}
\eeq%
therefore allows us to determine $\lambda_0$, which in our example 
is apparently close to 1. In our case with $|A_0|=|\ov A_0|$ (absence of
direct $C\!P$ violation) we have $|\lambda_0|=|q/p|$. With \eq{ekla} we find
\beq
\e_K \simeq \frac{1}{2} \lt[ 1 -\lambda_0  \rt] \; \simeq \;
    \frac{1}{2}
    \lt( 1 -\lt|\frac{q}{p}\rt| - i \, \imag \lambda_0 \rt) 
\label{ekqp}
\eeq
up to corrections of order $\e_K^2$. Remarkably, from the real and imaginary
part of $\e_K$ we infer two $C\!P$-violating quantities:
\begin{itemize}
\item[i)] the deviation of $|q/p|$ from 1 and
\item[ii)] the deviation of $ \imag \lambda_0$ from 0.
\end{itemize}
The first quantity is independent of the studied final state $f$ and codifies
\emph{$C\!P$ violation in mixing}. The second quantity, $\imag \lambda_f$, measures
$C\!P$ violation in the interference of mixing and decay or, in short,
\emph{mixing-induced $C\!P$ violation}\ in the decay $M\to f$.  

In the case of $D,B_d$ or $B_s$ mixing studies one tags the flavour at
some time $t=0$. The corresponding meson states are called $\ket{M(t)}$
and $\ket{\,\ov{\!M}(t)}$ and satisfy $\ket{M(t=0)}=\ket{M}$ and
$\ket{\,\ov{\!M}(t=0)}= \ket{\,\ov{\!M}}$. For $t>0$ these
time-dependent states are calculable superpositions of $\ket{M}$ and
$\ket{\,\ov{\!M}}$ and by observing the time-dependence of $M(t)\to f$
we can infer $\lambda_f$. The presently most prominent application of
this method is the precise determination of $\imag \lambda_f$ in the
decay $B_d \to J/\psi K_{\rm short}$ by the B factories BaBar and BELLE. Needless
to say that we will discuss this important topic in detail below.
  
%

While $C$, $P$, and $T$ are violated in nature, the combination
$C\!P\!T$ is a good symmetry. This \textit{CPT\ theorem} holds in any
local Poincar\'e-invariant quantum field theory \cite{lp}.  It implies
that particles and antiparticles have the same masses and total decay
widths.  When applied to our mixing problem characterised by $\Sigma$ in
\eq{defaij} the $C\!P\!T$ theorem enforces $\Sigma_{11}=\Sigma_{22}$.
However, while the $C\!P\!T$ theorem implies $\Gamma_{\rm tot}(M)=\Gamma_{\rm
  tot}(\ov M)$, one still has different time-integrated total decay
rates for tagged mesons, $\int_0^\infty \! dt\Gamma_{\rm tot}(M(t))\neq
\int_0^\infty \! dt \Gamma_{\rm tot}(\,\ov{\!M}(t))$.  This quantity is
sensitive to the ``arrow of time'' and the difference $\Gamma_{\rm
  tot}(M(t))- \Gamma_{\rm tot}(\,\ov{\!M}(t))$ measures $C\!P$ violation
rather than $C\!P\!T$ violation. Throughout my lectures I assume $C\!P\!T$
invariance and therefore identify $C\!P$ symmetry with $T$ symmetry.
Still, experiments have tested the $C\!P\!T$ theorem by probing
$\Sigma_{11}=\Sigma_{22}$ in \kkm.  We may speculate that Poincar\'e
invariance and $C\!P\!T$ symmetry are violated by the unknown dynamics of
quantum gravity. If we are lucky the size of $C\!P\!T$ violation scales
linearly in the inverse Planck Mass $M_{\rm Planck}$.  Interestingly,
today's accuracy of the $C\!P\!T$ test $\Sigma_{11}=\Sigma_{22}$ is roughly
$M_K/M_{\rm Planck}$.

\section{Second lecture: Time evolution}
\subsection{Time-dependent meson states}
In the Schr\"odinger picture, the time evolution of a quantum-mechanical
state $\ket{\psi}=\ket{\psi,t=0}$ is given by $\ket{\psi,t}={\cal
  U}(t,0)\ket{\psi} $, with the unitary time-evolution operator ${\cal
  U}(t,0)$. Consider first the case of a weakly-decaying charged meson
(i.e.\ $K^+$, $D^+$ or $B^+$), which cannot mix with other states. The
corresponding state at $t=0$, $\ket{M^+}$, will evolve into a
superposition of all states allowed by energy-momentum conservation.
This class of states consists of the original meson state $\ket{M^+}$
and all final states $\ket{f}$ into which $M^+$ can decay. Defining
\begin{eqnarray} 
   \ket{M^+ (t)} &=& \ket{M^+} \bra{M^+} {\cal U}(t,0) \ket{M^+}
\label{defmt}
\end{eqnarray}
we can write 
\begin{eqnarray} 
  {\cal U}(t,0) \ket{M^+} &=&  \ket{M^+ (t)} + 
            \sum_f \ket{f} \bra{f}  {\cal U}(t,0) \ket{M^+} .
\no  
\end{eqnarray}
In order to find $\ket{M^+ (t)}$ we take a shortcut, by 
employing the exponential decay law to deduce 
\begin{eqnarray}
   \ket{M^+ (t)} &=& e^{-i M_M t} e^{-\Gamma t/2} \ket{M^+}
  \label{explaw}  
\end{eqnarray}
in the meson rest frame.  The first term is the familiar time evolution
factor of a stable state with energy $E=M_M$. The second factor
involving the total width $\Gamma$ is understood by considering the
probability to find an undecayed meson at time $t$:
\begin{eqnarray}
\lt| \langle M^+ \ket{M^+ (t)}\rt|^2
   &=& e^{-\Gamma t} \no
\end{eqnarray}
Whenever I work in the Schr\"odinger picture I normalise the states as
$\langle M^+\ket{M^+}=1 $. Since $M_M-i \Gamma/2$ is
independent of $t$, we can compute it using the familiar covariant
formulation of quantum field theory. The optical theorem tells us that
$M_M$ and $-\Gamma/2$ are given by the real and imaginary parts of the
self-energy $\Sigma$ (depicted in the left diagram of \fig{fig:self}),
where%
\bea%
-i (2\pi)^4 \delta^{(4)}(\vec p\,{}^\prime-\vec p ) \Sigma &=&
\frac{\bra{M^+(\vec p\,{}^\prime)} S \ket{M^+ (\vec p)} }{2 M_M}%
\label{defsic}
\eea%
\begin{nfigure}{t}
\centering 
\includegraphics[scale=0.35]{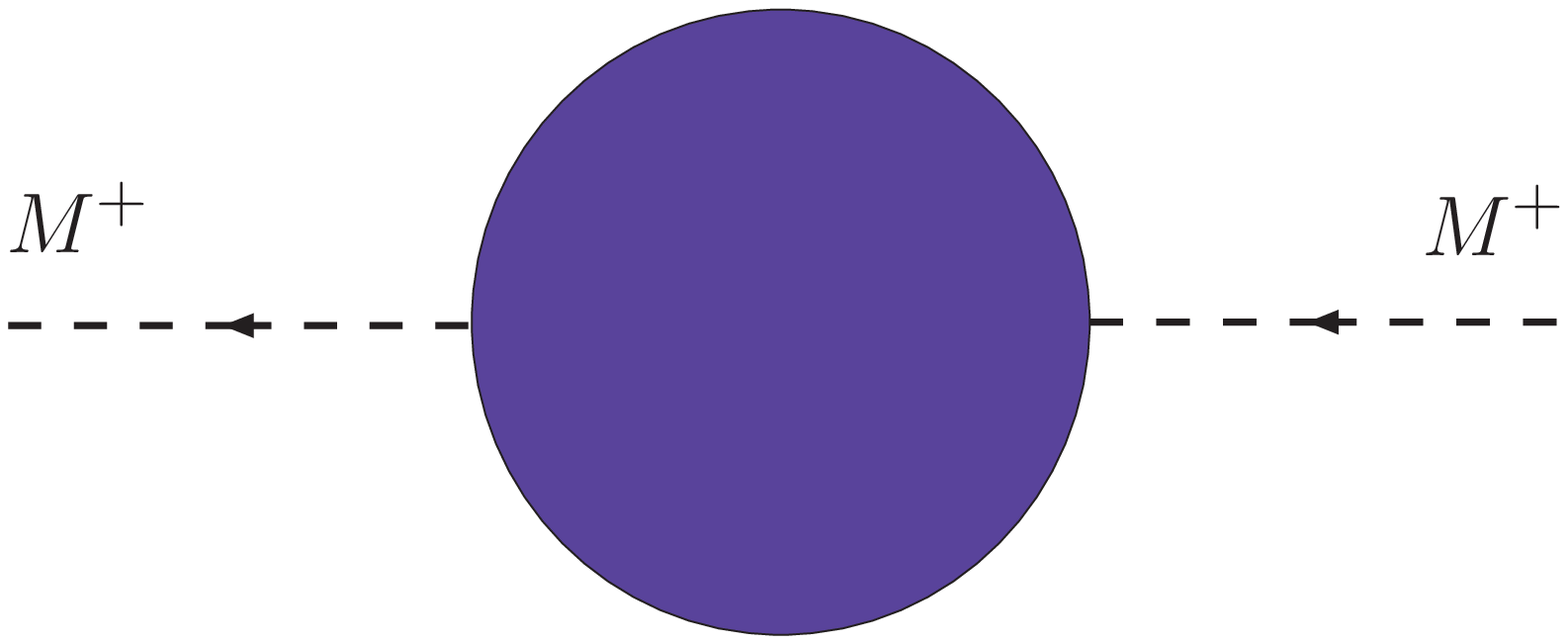} 
\hspace{2cm}
\includegraphics[scale=0.35]{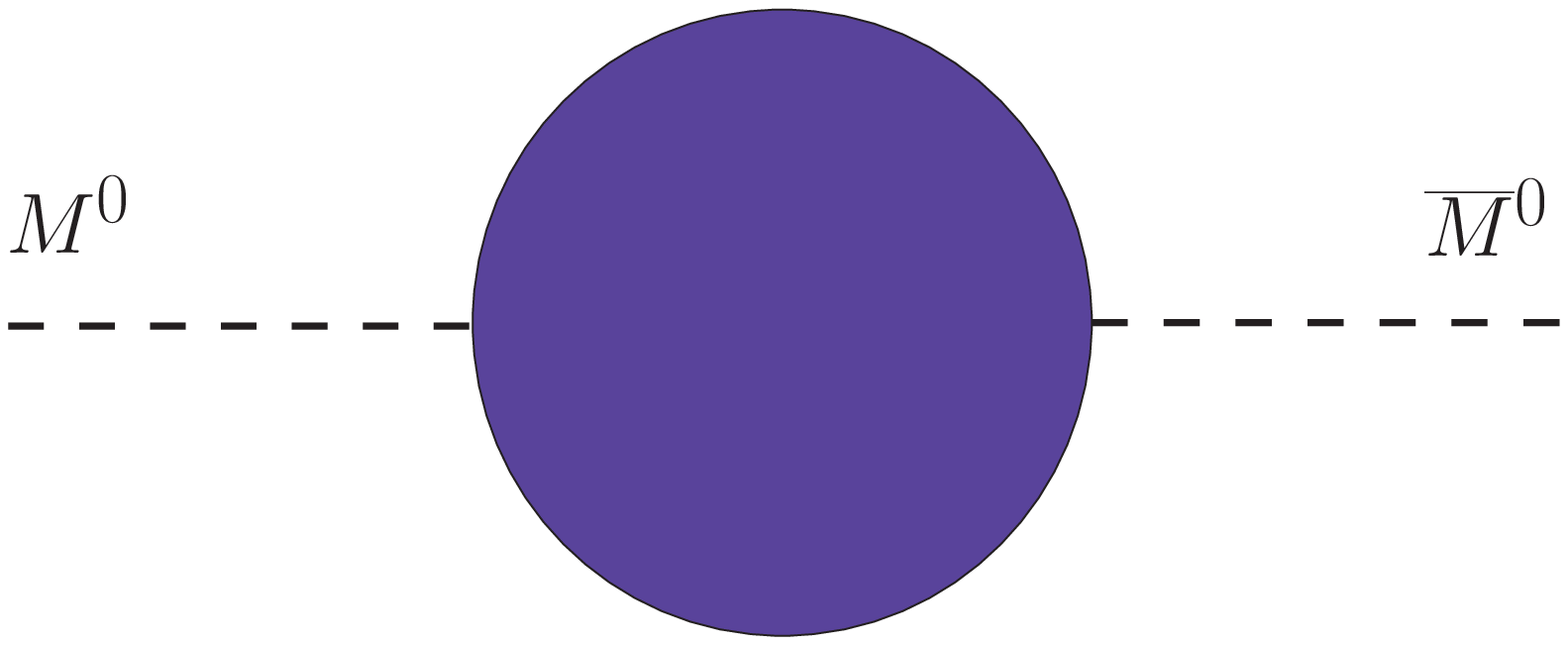} 
\caption{Left: generic self energy $\Sigma$ of a charged meson. Right: 
$M^0\!-\ov{\!M}{}^0$ mixing amplitude $\Sigma_{12}$.  
\label{fig:self}}
\end{nfigure}
(To be precise, the diagram in \fig{fig:self} corresponds to $2M_M
\Sigma$, so that $\Sigma=M_M-i\Gamma/2$ has mass dimension 1.)  From
\eq{explaw} we find
\begin{eqnarray}
 i \frac{d}{d\, t}  \ket{M^+ (t)} &=& 
    \Big( M_M-i \frac{\Gamma}{2}\, \Big) \ket{M^+ (t)} . \label{tevc} 
\end{eqnarray}
This equation can be generalised to a two-state system describing
neutral meson mixing: 
\begin{eqnarray}
 i \frac{d}{d\, t} \left( \!
\begin{array}{c}
\ds \ket{M (t)} \\[1mm]
\ds \ket{\,\ov{\!M} (t)}
\end{array}\! \right) &  = & 
\Sigma 
\left(\!
\begin{array}{c}
\ds \ket{M(t)} \\[1mm]
\ds \ket{\,\ov{\!M}(t)}
\end{array}\!\right) \label{schr}
\end{eqnarray}
where now $\Sigma$ is the $2\times 2$ matrix defined in \eq{defaij}.
Recall that any matrix can be written 
as the sum of a hermitian and an antihermitian matrix. We write
\begin{eqnarray}
\Sigma &=& M - i\, \frac{\Gamma}{2} \label{defmg}
\end{eqnarray}
with the \emph{mass matrix}\ $M=M^\dagger$ and the \emph{decay matrix}\
$\Gamma=\Gamma^\dagger$. Then
\begin{eqnarray}
  M_{12} &=& \frac{\Sigma_{12}+\Sigma_{21}^*}{2}, \qquad  \qquad  \qquad  
  \frac{\Gamma_{12}}{2} \;=\; i \, \frac{\Sigma_{12}-\Sigma_{21}^*}{2}. 
\label{absdisp}
\end{eqnarray}
The expressions on the RHS of \eq{absdisp} are called \emph{dispersive}\
and \emph{absorptive}\ parts of $\Sigma_{12}$, respectively. 
The right diagram in
\fig{fig:self} generically represents all contributions to
$\Sigma_{12}$. To compute $\Sigma_{12}$ we can certainly use
perturbation theory for the weak interaction (which to lowest order
amounts to the calculation of the box diagram in \fig{fig:boxes}), but
we must take into account the non-perturbative nature of the strong
binding forces.  The diagonal elements $M_{11}$ and $M_{22}$ are the
masses of $M$ and $\ov M$ and are generated from the quark mass terms in
$\cal L$ and from the binding energy of the strong interaction. However,
the off-diagonal elements $M_{12}= M_{21}^*$ and all elements of
$\Gamma$ stem from the weak interaction and are therefore tiny in
comparison with $M_{11}$ and $M_{22}$. The only reason why we can
experimentally access $M_{12}$ roots in the $C\!P\!T$ theorem: 
$C\!P\!T$ symmetry enforces
\begin{eqnarray}
  M_{11} &=& M_{22}, \qquad\qquad   
  \Gamma_{11} \; =\;  \Gamma_{22}, 
\label{cptmg}
\end{eqnarray}
so that the eigenvalues of $\Sigma$ are exactly degenerate for 
$\Sigma_{12}=\Sigma_{21}=0$. Even the smallest $\Sigma_{12}$ can lift
the degeneracy and can lead to large \mmm. 

With our shortcut we have avoided to prove that \eq{tevc} holds with
time-independent $M$ and $\Gamma$. In fact, \eq{tevc} and the inferred
exponential decay law in \eq{explaw} are not valid exactly, but receive
tiny (and phenomenologically irrelevant) corrections \cite{kha}. The same
statement is true for \eqsand{schr}{defmg}, a proper derivation of
\eq{schr} using time-dependent perturbation theory for the weak
interaction employs the so-called \emph{Wigner-Weisskopf approximation}\
\cite{wwa}. Corrections to this approximation have been addressed in 
Ref.~\cite{Chiu:1990cm} and are below the $10^{-10}$ level. 

We now proceed with the solution of our Schr\"odinger equation in
\eq{schr}.  \eq{defpq} means that the eigenvectors of $\Sigma$ in
\eq{defaij} are $(p,q)^T$ and $(p,-q)^T$.  That is, $\Sigma$ is
diagonalised as
\begin{eqnarray}
  Q^{-1} \Sigma \,Q &=& \lt( 
\begin{array}{cc} 
 M_L - i \Gamma_L/2 & 0 \\ 
 0 & M_H - i \Gamma_H/2 
\end{array} \rt)  \label{qsq}
\end{eqnarray}
with 
\begin{eqnarray}
  Q&=& \lt( 
\begin{array}{rr} 
p & p \\ q & - q \\
\end{array} \rt) \qquad \mbox{and} \qquad 
  Q^{-1} \;=\; \frac{1}{2 pq}  \lt( 
\begin{array}{rr} 
q &  p \\ q & - p \\
\end{array} \rt)   .\label{defq}
\end{eqnarray}
The ansatz in \eq{defq} works because $\Sigma_{11}=\Sigma_{22}$.  The
mass eigenstates $\ket{M_{L,H}(t)}$ obey an exponential decay law as
$\ket{M^+ (t)}$ in \eq{explaw} with $(M_M,\Gamma)$ replaced by
$(M_{L,H},\Gamma_{L,H})$.  Transforming back to the flavour basis gives
%
\bea%
\lt( \!\begin{array}{c} 
\ket{M (t)} \\[2mm] 
\ket{\,\ov{\!M} (t)} 
\end{array}\! \rt) 
&=& Q \, \lt(
\begin{array}{cc} 
\ds e^{-i M_Lt -  \Gamma_L t/2} & \ds 0 \\[2mm] 
\ds  0 & e^{-i M_H t - \Gamma_H t/2} 
\end{array} \rt)\, Q^{-1} 
\lt( \! \begin{array}{c} 
\ket{M } \\[2mm] 
\ket{\,\ov{\!M} } 
\end{array} \!\rt)
\label{tes}
\eea%
I adopt the following definitions for the average mass and width and
the mass and width differences of the mass eigenstates:
\begin{equation}
\begin{array}{rclrcl}
m & = & \displaystyle \frac{ M_H + M_L}{2} =  M_{11} = M_{22} \,, \qquad &
  \Gamma & = & \displaystyle \frac{\Gamma_L + \Gamma_H}2 = 
  \Gamma_{11} = \Gamma_{22} \,,
  \\[6pt]
\dm & = & M_H - M_L \,, & \dg & = & \Gamma_L - \Gamma_H \,.
\end{array}\label{mg}
\end{equation}
Note that $\dm $ is positive by definition while $\dg$ can have either sign.
Experimentally the sign of $\dg$ is only known for Kaons and my sign
convention in \eq{mg} corresponds to $\dg_K >0$. The Standard-Model prediction
for $\dg_{B_d}$ and  $\dg_{B_s}$ is also positive, while no reliable
prediction is possible for the sign of $\dg_D$.  
The matrix appearing in \eq{tes} can be compactly written as 
\begin{equation}
 Q \, \lt( 
\begin{array}{cc} 
\ds e^{-i M_L t- \Gamma_L t/2} & \ds 0 \\[2mm] 
\ds  0 & e^{-i M_H t- \Gamma_H t/2} 
\end{array} \rt) \, Q^{-1} \;=\; 
\lt( 
\begin{array}{rr}
\ds g_+(t)              &\ds \frac{q}{p} g_-(t) \\[2mm]  
\ds \frac{p}{q} g_-(t)  &\ds g_+ (t)  
\end{array} \rt) \label{qlhq}
\end{equation}
with%
\bea%
 g_+ (t) &=& e^{-i m t} \, e^{-\Gamma t/2} \lt[ \phantom{-} \cosh\frac{\dg \,
  t}{4}\, \cos\frac{\dm\, t}{2} - i \sinh\frac{\dg \, t}{4}\, \sin \frac{\dm
  \, t}{2} \, \rt] , \nn g_- (t) &=& e^{-i m t} \, e^{-\Gamma t/2} \lt[-
\sinh\frac{\dg \, t}{4}\, \cos \frac{\dm \, t}{2} + i \cosh\frac{\dg \,
  t}{4}\, \sin \frac{\dm \, t}{2} \, \rt] .
\label{gpgm}
\eea%
Inserting \eq{qlhq} into \eq{tes} gives us a transparent picture of the 
\mm\ oscillations: 
\bea%
\ket{M (t)} &=& \phantom{\frac{p}{q}\,}
  g_+ (t)\, \ket{M} + \frac{q}{p}\, g_- (t)\, \ket{\,\ov{\!M}} \,, \nn
\ket{\,\ov{\!M} (t)} &=& \frac{p}{q}\, g_- (t)\, \ket{M}
  + \phantom{\frac{q}{p}\,} g_+(t)\, \ket{\,\ov{\!M}} \,,
  \label{tgg}
  \eea%
  We verify $g_+(0)=1$ and $g_-(0)=0$ and find that $g_\pm (t)$ has no
  zeros for $t>0$ if $\dg\neq 0$. Hence an initially produced
  $M$ will never turn into a pure $\,\ov{\!M}$ or back into a pure $M$.
We will frequently encounter the combinations%
  \bea%
  | g_\pm (t) |^2 & = & \frac{e^{- \Gamma t}}{2} \lt[ \phantom{-} \cosh
  \frac{\Delta \Gamma \, t}{2} \pm \cos\lt( \dm\, t \rt) \rt] , \nn
  g_+^* (t)\, g_- (t) & = & \frac{e^{- \Gamma t}}{2} \lt[ - \sinh
  \frac{\Delta \Gamma \, t}{2} + i \sin \lt( \dm\, t \rt) \rt] .
\label{gpgms}
\eea%
\boldmath
\subsection{\dm, \dg\ and CP violation in mixing}
\unboldmath
We still need to solve our eigenvalue problem. The secular equation for the 
two eigenvalues $\sigma_{L,H}=M_{L,H}-i\Gamma_{L,H}/2$ of $\Sigma$ is 
$(\Sigma_{11}-\sigma_{L,H})^2-\Sigma_{12}\Sigma_{21} =0$. The two solutions of 
this equation therefore satisfy %
\bea%
\lt(\sigma_H -\sigma_L\rt)^2 &=& 4\,\Sigma_{12}\Sigma_{21} \no 
\eea%
or 
\bea%
( \dm + i \frac{\dg }{2} )^2 
\;= \; 4\,
     \lt( M_{12}   -  i \frac{\Gamma_{12}}{2} \rt) 
     \lt( M_{12}^* -  i \frac{\Gamma_{12}^*}{2} \rt). 
\eea%
Taking real and imaginary part of this equation leads us to 
\bea%
\lt( \dm \rt)^2 - \frac{1}{4} \lt( \dg \rt)^2
  &=& 4  \lt| M_{12} \rt|^2 -  \lt| \Gamma_{12} \rt|^2 \,,
  \label{mgqp:a} \\[2mm]
\dm\, \dg &=& -4\, \real ( M_{12} \Gamma_{12}^* ) \,,
  \label{mgqp:b} 
\eea%
Further \eq{qsq} implies $[Q^{-1} \Sigma Q]_{12}=[Q^{-1} \Sigma
Q]_{21}=0$, which determines 
\bea%
\frac{q}{p} &=& - \frac{\dm + i \, \dg/2}{2 M_{12} -i\, \Gamma_{12} }
  = - \frac{2 M_{12}^* -i\, \Gamma_{12}^*}{\dm + i \, \dg/2} \,.
\label{mgqp:c}
\eea%
(There is also a second solution with the opposite sign, which, however, is
eliminated by imposing $\dm >0$.)  For the simplification of
\eqsto{mgqp:a}{mgqp:c} it is useful to identify the physical quantities
of the mixing problem in \eqsand{schr}{defmg}. In quantum mechanics we
can always multiply either $\ket{M}$ or $\ket{\,\ov{\!M}}$ by an
arbitrary phase factor without changing the physics. This will change
the phases of $M_{12}$, $\Gamma_{12}$ and $q/p$, none of which can
therefore have any physical meaning. The three physical quantities of
\mmm\ are \bea%
&& |M_{12}|, \qquad\qquad |\Gamma_{12}|, \qquad\quad \mbox{and} \quad
\phi = \arg \left( -\frac{M_{12}}{\Gamma_{12}} \right).
        \label{defphi} 
\eea%
\eq{mgqp:b} then reads 
\bea%
\dm\, \dg &=& 4\, |M_{12}|  |\Gamma_{12}| \cos\phi .
\label{mgqp:d}
\eea%
We can easily solve \eqsand{mgqp:a}{mgqp:d} to express $\dm$ and $\dg$, which
we want to measure by studying meson time evolutions, in terms of the
theoretical quantities $|M_{12}|$, $|\Gamma_{12}|$ and $\phi$.  We recognise
that the phase $\phi$ is responsible for $C\!P$ violation in mixing introduced
after \eq{ekqp}: By multiplying the two expression for $q/p$ in \eq{mgqp:c}
with each other we find
\bea%
\lt( \frac{q}{p}\rt)^2 
 &=& \frac{2 M_{12}^* -i\, \Gamma_{12}^*}{2 M_{12} -i\, \Gamma_{12} }
 \; =\;  \frac{M_{12}^*}{M_{12}} \, 
      \frac{\ds 1 + i \lt|\frac{\Gamma_{12}}{2 M_{12}}\rt| e^{ i  \phi\;\,}}
           {\ds 1 + i \lt|\frac{\Gamma_{12}}{2 M_{12}}\rt| e^{ -i \phi}}
\label{mgqp:e} .
\eea%
We immediately verify from this expression that $\phi\neq 0,\pi$ indeed
implies $|q/p|\neq 1$, which defines $C\!P$ violation in mixing.

Interestingly, $C\!P$ violation in mixing is small (if quantified in terms
of $|q/p|-1$) for the $K$, $B_d$ and $B_s$ systems. For \ddm\ this is
most likely also the case, but the experimental data are not accurate
enough at present.  In the case of \kkm\ we have established this
phenomenon in \eq{ekqp} from the measured value of $\real \e_K$ in
\eq{ekexp}. In the \bb\ systems the line of arguments is as follows:
Experimentally we know $\dm \gg \dg$ and theoretically $|\Gamma_{12}|\ll
\dm $ is firmly established from a SM calculation, since the possible
impact of new physics on $|\Gamma_{12}|$ is small. Then
\eqsand{mgqp:a}{mgqp:d} imply $\dm \approx 2 |M_{12}|$ and therefore
$|\Gamma_{12}|\ll | M_{12} |$, so that the second term in the numerator
and denominator of \eq{mgqp:e} is small, irrespective of the value of
$\phi$.  Thus $|q/p|\simeq 1$ for $B_d$ and $B_s$ mesons. It is useful
to define the quantity $a$ through
\begin{eqnarray}
\lt| \frac{q}{p} \rt|^2 &=& 1 -a . \label{defa} 
\end{eqnarray}
For the $K$, $B_d$ and $B_s$ systems we know that $a$ is small. 
By expanding $(q/p)^2$ in \eq{mgqp:e} in terms of $\phi$ or 
$\Gamma_{12}/M_{12}$ we find
\begin{eqnarray}
 a &=&   \frac{4 |\Gamma_{12}|\, |M_{12}|}{
               4 |M_{12}|^2 + |\Gamma_{12}|^2} \, \phi 
          + {\cal O} (\phi^2), 
     \qquad\qquad\qquad\qquad\qquad\mbox{for \kkm} \label{agmk} \\ 
 a &=& \imag \frac{\Gamma_{12}}{M_{12}} + 
         {\cal O} \lt(\Big(  \imag \frac{\Gamma_{12}}{M_{12}} \Big)^2   \rt)
        \; = \;  \lt| \frac{\Gamma_{12}}{M_{12}} \rt| \sin \phi \,, 
      \qquad\quad\mbox{for \bbm}.   
    \label{agmb} 
\end{eqnarray}
With this result it is straightforward to solve \eqsand{mgqp:a}{mgqp:d} 
for $\dm$ and $\dg$. Incidentally, in both cases we have 
\bea%
\dm &\simeq & 2\, |M_{12}| ,
  \label{mgsol:a} \\
\dg &\simeq & 2\, |\Gamma_{12}| \cos \phi  .
  \label{mgsol:b}
\eea%
which holds up to corrections of order $\phi^2$ for Kaons and of order 
$|\Gamma_{12}/M_{12}|^2$ for $B$ mesons. Of course, in the former case
one can also replace $\cos \phi$ by 1. Importantly, in $B$ physics 
one deduces from \eq{mgqp:c} that
\bea%
  \frac{q}{p} &=& - \frac{M_{12}^*}{|M_{12}|} 
     \lt[ 1 + {\cal O} (a) \rt] \label{phqp} .
\eea%
That is, the phase of $-q/p$ is essentially given by the phase of 
the \bbd\ or \bbs\ box diagram in \fig{fig:boxes}. Since \bbm\ is
dominated by the box diagram with internal tops we readily infer 
\bea%
     \frac{q}{p} &=& -\frac{V_{tb}^* V_{tq}}{V_{tb} V_{tq}^*}
      \; =\; - \exp[ i \arg \lt( V_{tb}^* V_{tq} \rt)^2] \qquad\qquad 
   \mbox{for \bbmq\ with $q=d,s$} \label{qpb}   
\eea%
up to tiny corrections of order $a$.

\subsection{Time-dependent decay rates}\label{sec:time} 
Flavour factories are $e^+e^-$ colliders whose CMS energy matches the
mass of an excited quarkonium state which predominantly decays into
$(M,\Mbar)$ pairs. Running on the $\psi(3770)$, $\Upsilon(4S)$ or
$\Upsilon(5S)$ resonances, one copiously produces $(D, \Dbar)$, $(B_d,
\Bbar_d)$ or $(B_s,\Bbar_s)$ mesons. The $(M,\Mbar)$ pairs are in an
entangled quantum-mechanical state until the decay of one of the mesons
is observed. If the decay mode $ M\to f$ is allowed while $\Mbar \to f$
is forbidden one calls $ M\to f$ a \emph{flavour-specific}\ mode or a
\emph{tagging mode}.  The most prominent examples are the semileptonic
decays $M \to X \ell^+ \nu_\ell$. For the discovery of \bbms\ the
flavour-specific mode $B_s \to D_s^- \pi^+$ has played an important role
\cite{cdf}. A flavour-specific decay tags the decaying meson as either
$M$ or $\Mbar$. The Einstein-Podolsky-Rosen effect then ensures that the
other meson is an $\Mbar$ or $M$, respectively. The time of the flavour
tagging ``starts the clock'', i.e.\ defines $t=0$ in \eqsand{gpgm}{tgg}.
This method is called \emph{opposite-side tagging}. In hadron colliders 
pairs of different hadrons can be produced, e.g.\ a $B_s$ can be
produced together with a $B^-$ or $\Lambda_b$ plus several lighter
hadrons. Still, at the quark level $(\ov b,b)$ pairs are produced, so
that the flavour tagging works as well. As an additional possibility, 
hadron colliders permit \emph{same-side tagging}, where the flavour is
determined at the time of the hadronisation process: When, say, a
$b$-quark hadronises into a $\Bbar$ meson several pions and Kaons are
produced as well. The charges of these light mesons are correlated 
with the charge of the light valence quark, which in the case of the   
$\Bbar$ meson is an anti-$d$ quark. 

The time-dependent decay rate of a meson tagged at $t=0$ as $M$ is
defined as%
\beq%
\gtf = \frac{1}{N_M}\, \frac{d\, N(M (t) \to f)}{d\, t} \,,
\label{defgtf}
\eeq%
where $d\, N(M (t) \to f)$ denotes the number of decays into the final
state $f$ occurring within the time interval between $t$ and $t+d\, t$.
$N_M$ is the total number of $M$'s produced at time $t=0$. An analogous
definition holds for \gbtf.  One has%
\beq%
\gtf = {\cal N}_f \lt| \bra f S \ket{M (t)}  \rt|^2 , 
  \qquad \gbtf =
{\cal N}_f \lt| \bra f S \ket{\Mbar (t)} \rt|^2
\label{gtfaf}
\eeq%
with the time-independent normalisation factor $ {\cal N}_f$ comprising
the result of the phase-space integration.  It is straightforward to
calculate $\gtf$ and $\gbtf$ in terms of $A_f$ and $\ov A_f$ defined in
\eq{defaf}, we just need to insert $\ket{M(t)}$ and $\ket{\Mbar(t)}$
from \eq{tgg} into \eq{gtfaf}. Trading $\ov A_f$ for $\lambda_f$ (see
\eq{deflaf}) and $a$ (see \eq{defa}) and making use of \eq{gpgms} we
find the desired formulae:%
\bea%
\gtf &=& {\cal N}_f \, | A_f |^2 \, e^{-\Gamma t}\, \Bigg\{ \frac{1 +
  \lt| \lambda_f \rt|^2}2\, \cosh \frac{\dg \, t}{2} +
\frac{ 1 - \lt| \lambda_f \rt|^2}2\, \cos ( \dm \, t )  \no \\*
&& \qquad \qquad \qquad - \real \lambda_f \, \sinh \frac{\dg \, t}{2} -
\imag \lambda_f \, \sin \lt( \dm \, t \rt) \Bigg\} \,,
\label{gtfres} \\
\gbtf &=& {\cal N}_f \, | A_f |^2 \, \frac{1}{1-a} \,
  e^{-\Gamma t}\, \Bigg\{ \frac{1 + \lt| \lambda_f \rt|^2}2\,
    \cosh \frac{\dg \, t}{2}
  - \frac{1 - \lt| \lambda_f \rt|^2}2\, \cos ( \dm \, t ) \no \\*
&&  \qquad \qquad \qquad
    - \real \lambda_f \, \sinh \frac{\dg \, t}{2}
    + \imag \lambda_f \, \sin ( \dm \, t ) \Bigg\} \,.
   \label{gbtfres}
\eea%
Often we want to compare these decay modes with the corresponding 
decays into the final state which is CP-conjugate with respect to $f$.  
For states $f$ with two or more particles we define%
\beq%
\ket{\ov{f}} = CP\, \ket{f} \,, \label{defcpf}%
\eeq%
while for the initial one-particle states we have defined $C\!P$ in
\eq{defcpm}.  For example, for $f=D_s^- \pi^+$ the $C\!P$-conjugate state
is $\ov{f}=D_s^+\pi^-$. Whenever we discuss $C\!P$ (or any other
discrete transformation) in decay processes, we apply the transformation
in the rest frame of the decaying meson. The transformation in
\eq{defcpf} is understood to reverse the signs of three-momenta as in
\eq{defcpm}. For two-body final states, which are our prime focus, we
can rotate this mirror-reflected state by 180$^\circ$, so that the
three-momenta of the rotated $C\!P$-transformed state coincide with
those of the original state.  This procedure is usually implicitly
understood when people discuss decays into $C\!P$ eigenstates composed
of two distinct particles, such as $K\to \pi^+\pi^-$. For a $C\!P$
eigenstate $f_{\rm CP}$ \eqsand{defcpeig}{defcpf} imply 
$\ket{\ov{f}_{\rm CP}}=\eta_{f_{CP}} \ket{f_{\rm CP}}$. 

In the $M(t) \to \ov f$ decay rates   it is advantageous to keep 
$\ov{A}_{\ov f}$ while trading $A_{\ov f}$ for $\lambda_{\ov f}$:%
\bea%
\gtfb &=&  {\cal N}_f \lt| \ov{A}_{\ov{f}} \rt|^2 e^{-\Gamma t}\,
  ( 1 - a) \, \Bigg\{ \frac{1 +
  | \lambda_{\ov{f}} |^{-2}}{2}\, \cosh \frac{\dg \, t}{2}
  - \frac{ 1 - | \lambda_{\ov{f}} |^{-2}}{2}\, \cos ( \dm \, t ) \no \\*
&& \qquad \qquad \qquad
  - \real \frac{1}{\lambda_{\ov{f}}}\, \sinh \frac{\dg \, t}{2} \,
  + \imag \frac{1}{\lambda_{\ov{f}}}\, \sin (\dm \, t) \Bigg\} \,,
\label{gtfbres} \\[2pt]
\gbtfb &=& {\cal N}_f \lt| \ov{A}_{\ov{f}} \rt|^2 e^{-\Gamma t}\,
  \Bigg\{ \frac{1 + | \lambda_{\ov{f}} |^{-2}}2\,
    \cosh \frac{\dg \, t}{2}
  + \frac{1 - | \lambda_{\ov{f}} |^{-2}}2\, \cos ( \dm \, t ) \no\\*
&& \qquad \qquad \qquad
  - \real \frac{1}{\lambda_{\ov{f}}}\, \sinh \frac{\dg \, t}{2}
  - \imag \frac{1}{\lambda_{\ov{f}}}\, \sin ( \dm \, t ) \Bigg\} \,.
   \label{gbtfbres}
\eea%

\eqsto{gtfres}{gbtfres} and \eqsto{gtfbres}{gbtfbres} are our master
formulae to calculate any time-dependent decay rate of interest.  We
discuss two important applications here. The first one is the time
dependence of a flavour-specific decay, which satisfies $\ov A_f=A_{\ov
  f}=\lambda_f=1/\lambda_{\ov f}=0$.  In addition we consider a decay
mode with $|\ov A_{\ov f}|=|A_f|$, that is without direct CP violation.
Semileptonic decays satisfy both conditions. Our master formulae become
very simple for this case. Defining the \emph{mixing asymmetry},%
\beq%
{\cal A}_0 (t) = \frac{\gtf - \gtfb}{\gtf + \gtfb } \,,
\label{defa0}
\eeq%
one finds to order $a$:%
\beq%
{\cal A}_0 (t) = \frac{\cos ( \dm\, t ) }{\cosh (\dg \,t/2)}
  + \frac{a }{2}
  \lt[1- \frac{\cos^2 ( \dm\, t )}{\cosh^2 (\dg \, t/2)}  \rt] .
\label{resa0}
\eeq%
Note that ${\cal A}_0 (t)$ is not a $C\!P$ asymmetry. Instead $\gtf\
\propto |\bra{M} M(t)\rangle|^2$ is proportional to the probability that
an ``unmixed'' $M$ decays to $f$ at time $t$, while $\gtfb\ \propto
|\bra{\ov M} M(t)\rangle|^2$ is the corresponding probability for the
process $M\to\ov M \to f$. The asymmetry ${\cal A}_0 (t)$ is often
employed to measure $\dm$. In the ARGUS discovery of \bbmd\ \cite{argus}
no time-dependence was observed. Instead so-called like-sign dilepton
events were observed in semileptonic $(B_d,\Bbar_d)$ decays, meaning
that one of the two mesons must have mixed. By counting these events and
comparing the number with the number of opposite-sign dilepton events
one can infer the quantity $x=\dm/\Gamma$. The corresponding formula
can be found by integrating our master formulae over $t$. 

The \emph{CP asymmetry in flavour-specific decays} (often called
\emph{semileptonic CP asymmetry}) reads%
\bea%
a_{\rm fs} & \equiv & \frac{\gbtf - \gtfb}{\gbtf + \gtfb} \; =\; \frac{
  1-(1-a)^2 }{1+(1-a)^2} \;=\;a + {\cal O} (a^2) .
\label{afst} 
\eea%
Define the
untagged decay rate
\begin{eqnarray}
\guntf &=& \gbtf + \gtf         \label{guntf}
\end{eqnarray}
to find:
\begin{eqnarray}
a_{\rm fs, unt}(t) &=&
   \frac{\guntf - \guntfb}{\guntf + \guntfb}
\;= \;
        \frac{a_{\rm fs}}{2} - \frac{a_{\rm fs}}{2} \,
        \frac{\cos ({ \dm} \, t)}{\cosh ({ \dg} t/2) }
        . \,  \label{fsun}
\end{eqnarray}
Hence no tagging is needed to measure $a_{\rm fs}$!  We observe that we can
determine the three physical quantities characterising \mmm, $|M_{12}|$,
$|\Gamma_{12}|$ and $a$, by measuring \dm, \dg\ and $a_{\rm fs}$. At present
all three quantities are only measured for \kkm! Also the semileptonic $C\!P$
asymmetry of $B$ mesons can be measured without observing any time dependence.
In the spirit of ARGUS we can compare the number of positively-charged
like-sign dilepton pairs with the number of negatively-charged ones. Such
measurements are performed at the B factories and the Tevatron, but no
non-zero semileptonic $C\!P$ asymmetry has been established by now.

Amusingly, the oscillations drop out from the tagged quantity
in \eq{afst}, while they persist in \eq{fsun}. In most applications one
can neglect the tiny $a$ in \eqsto{gtfres}{gbtfres} and
\eqsto{gtfbres}{gbtfbres}. Then we realise that in the untagged rates,
obtained by adding \eqsand{gtfres}{gbtfres} or
\eqsand{gtfbres}{gbtfbres}, the terms involving $\cos(\dm t)$ and $\sin
(\dm t)$ vanish. 

The second application of our master formulae are decays into CP
eigenstates, $M\to f_{\rm CP}$. The time-dependent $C\!P$ asymmetry is
\beq
a_{f_{\rm CP}}(t) 
  = \frac{ \gbtfcp - \gtfcp }{ \gbtfcp + \gtfcp } \,. \label{defacp}
\eeq
Using \eq{gtfres} and \eq{gbtfres} one finds
\beq
a_{f_{\rm CP}}(t) = - \frac{A_{CP}^{\rm dir} \cos ( \dm  \, t ) +
       A_{CP}^{\rm mix} \sin ( \dm \, t )}{
       \cosh (\dg \, t/ 2) +
       A_{\dg} \sinh  (\dg \, t / 2) }
    + {\cal O} ( a ) \,,
    \label{acp}%
\eeq%
with (for $f=f_{\rm CP}$)%
\beq A_{CP}^{\rm dir} = \frac{1- \lt| \lambda_f \rt|^2}{1+ \lt|
  \lambda_f \rt|^2} \,, \qquad 
A_{CP}^{\rm mix} = - \frac{2\, \imag
  \lambda_f}{1+ \lt| \lambda_f \rt|^2} \,, 
\qquad
A_{\dg} = -
\frac{2\, \real \lambda_f}{1+ \lt| \lambda_f \rt|^2} .
    \label{dirmix}
\eeq%
Note that $|A_{CP}^{\rm dir}|^2 + |A_{CP}^{\rm mix}|^2 +
|A_{\Delta\Gamma}|^2 = 1$. Experimentally one can track the
time-dependence of $a_f(t)$ and read off the coefficients of $\cos (
\dm \, t )$ and $\sin ( \dm \, t )$, so that one can determine
$|\lambda_f|$ and $\imag \lambda_f$.  When studying decay amplitudes we
can treat the weak interaction perturbatively by drawing quark-level
Feynman diagrams involving the exchange of W-bosons.  While we cannot
fully compute those diagrams, because we cannot estimate how the quarks
are ``dressed'' by the strong interaction, we can still assess the
CP-violating phases by identifying the CKM elements in the diagrams.
Decays in which all contributing Feynman diagrams carry the same
CP-violating phase are called \emph{golden modes}. These modes satisfy
$|A_f|=|\ov A_{\ov f}|$, so that there is no direct CP violation.  In a
golden $M\to f_{\rm CP}$ decay this means $|\lambda_{f_{\rm CP}}|=1$ and
in \eqsand{acp}{dirmix} we have $A_{CP}^{\rm dir}=0$ and%
\beq%
A_{CP}^{\rm mix} = \imag \lambda_{f_{\rm CP}} . \label{acpgold} \eeq%
Moreover the phase of $\ov A_{f_{\rm CP}}/A_{f_{\rm CP}}$ is trivially
read off from the phase of the CKM elements. In $B$ physics, where we
also know the phase of $q/p$ from \eq{qpb}, we can therefore directly
relate the measured $ \imag \lambda_{f_{\rm CP}}$ to phases of CKM
elements, if  $M\to f_{\rm CP}$ is golden.

\section{Third lecture: Linking quarks to mesons}  
\subsection{The Cabibbo-Kobayashi-Maskawa matrix}
We have encountered the CKM matrix $V$ in \eq{defv}.  A unitary
$3\times3 $ matrix can be parameterised by three angles and six complex
phases. With the rephasings in \eqsand{fph}{vph} we can eliminate five
phases from $V$ leaving us with one physical $C\!P$-violating phase. In the
parameterisation favoured by the Particle Data Book one has
\begin{equation}
    V = \left( \begin{array}{ccc}
                c_{12}c_{13} & s_{12}c_{13} & s_{13}e^{-i\delta_{13}} \\
                -s_{12}c_{23}-c_{12}s_{23}s_{13}e^{i\delta_{13}} &
                 c_{12}c_{23}-s_{12}s_{23}s_{13}e^{i\delta_{13}} &
                s_{23}c_{13} \\
                 s_{12}s_{23}-c_{12}c_{23}s_{13}e^{i\delta_{13}} &
                -c_{12}s_{23}-s_{12}c_{23}s_{13}e^{i\delta_{13}} &
                c_{23}c_{13}
        \end{array} \right),
    \label{VPDG}
\end{equation}
where $c_{ij}=\cos\theta_{ij}$ and $s_{ij}=\sin\theta_{ij}$.
The real angles~$\theta_{ij}$ may be chosen so that
$0\le\theta_{ij}\le\pi/2$, and the phase $\delta_{13}$ so that
$-\pi < \delta_{13} \leq \pi$. 
For the discussion of CKM metrology it is useful to introduce the Wolfenstein
parameterisation \cite{w}
\begin{equation}
    V = \left( \begin{array}{ccc}
                1-\frac{1}{2}\lambda^2 & \lambda & A\lambda^3(\rho-i\eta) \\
                -\lambda & 1-\frac{1}{2}\lambda^2 & A\lambda^2 \\
                 A\lambda^3(1-\rho-i\eta) & -A\lambda^2 & 1
        \end{array} \right) + O(\lambda^4)\, ,
\label{wolf}
\end{equation} 
which is an expansion in terms of the small parameter $ \lambda=0.22$.
The remaining three parameters $A$, $\rho$ and $\eta$ are a bit smaller
than 1. The Wolfenstein parameterisation nicely reveals the hierarchical
structure of the CKM matrix, with diagonal elements of order 1 and
smallest elements in the upper right and lower left corners.  We can now
understand why the prediction of $m_c$ from $\dm_K$ in 1974 was
successful: Any contribution involving the top quark (at that time
unknown and unimagined by the authors of Ref.~\cite{gl}) to the upper
left diagram in \fig{fig:boxes} is highly suppressed by small CKM
elements, since $|V_{td} V_{ts}|\simeq \lambda^5$, while $|V_{cd}
V_{cs}|\simeq |V_{ud} V_{us}| \simeq \sin\theta_c\simeq \lambda $.
Further the upper left $2\times 2$ submatrix, the Cabibbo matrix, is
almost unitary and involves only a single parameter, the Cabibbo angle
$\theta_c$ with $V_{ud} \simeq V_{cs} \simeq \cos \theta_c$ and $V_{us}
\simeq -V_{cd} \simeq \lambda $. Therefore the two new elements $V_{cd}$
and $V_{cs}$ predicted in Ref.~\cite{gim} were completely fixed in terms
of the known $\theta_c$. In the Wolfenstein approximation only $V_{ub}$
and $V_{td}$ have a complex phase and $C\!P$ violation is characterised
by $\eta\neq 0$.

Any unitary $3\times 3$ matrix satisfies  
\begin{eqnarray}
   V_{1j}^* V_{1k} + V_{2j}^* V_{2k} + V_{3j}^* V_{3k} &=& \delta_{jk}
   \label{un1} \\ 
\mbox{and}\qquad\qquad
    V_{j1}^* V_{k1} + V_{j2}^* V_{k2} + V_{j3}^* V_{k3} &=& \delta_{jk}
 \label{un2} .
\end{eqnarray}
If we choose $j\neq k $ the three terms add to zero. We can depict the
relations in \eqsand{un1}{un2} as triangles in the complex plane, e.g.\
for \eq{un1} the three corners are located at $0$, $ V_{1j}^* V_{1k}$
and $-V_{2j}^* V_{2k}$. The three sides can be associated with the three
terms summing to zero.  The area of all six triangles is the same and
given by $J/2$, where $J$ is the \emph{Jarlskog invariant} \cite{j}
\begin{eqnarray} 
 J &\equiv & \imag \lt[ V_{td}^* V_{tb} V_{ub}^* V_{ud} \rt]  
 \; =\, c_{12} c_{23} c_{13}^2 s_{12} s_{23} s_{13} \sin\delta_{13} 
 \; \simeq \; A^2\lambda^6\eta .
\end{eqnarray}
Here the third expression refers to the exact parameterisation of
\eq{VPDG} and the last result uses the Wolfenstein approximation.  Four
of the six \emph{unitarity triangles} are squashed, the three sides are
similar only for the choice $(j,k)=(3,1)$. Moreover, within the
Wolfenstein approximation the shapes of the triangles corresponding to
\eqsand{un1}{un2} are equal for $(j,k)=(3,1)$. Applying the phase
transformations of \eqsand{fph}{vph} rotates the unitarity triangles in
the complex plane, but leaves their shape fixed.  Seeking a definition
of a rephasing-invariant unitarity triangle with a physical meaning we
divide \eq{un1} (for $(j,k)=(3,1)$) by $V_{23}^* V_{21}=V_{cb}^*
V_{cd}$ to arrive at
\begin{eqnarray} 
  \frac{V_{ub}^* V_{ud}}{V_{cb}^* V_{cd}} +     
  \frac{V_{tb}^* V_{td}}{V_{cb}^* V_{cd}} +  1 &=& 0
\label{sut}
\end{eqnarray}
When people speak of ``the'' unitarity triangle they refer to the
rescaled triangle defined by \eq{sut}. Since its baseline coincides with
the interval $[0,1]$ of the real axis, the unitarity triangle is
completely determined by the location of its apex $(\ov \rho,\ov \eta)$,
where 
\begin{eqnarray}
\ov \rho + i \ov \eta \equiv - \frac{V_{ub}^* V_{ud}}{V_{cb}^* V_{cd}} 
 \label{defre} .
\end{eqnarray}
Inserting \eq{wolf} into \eq{defre} one realises that $(\ov \rho,\ov
\eta)= (\rho,\eta)$ within the Wolfenstein approximation, which here is
good to an accuracy of 3\%. The unitarity triangle is depicted in 
\fig{fig:ut}.
\begin{nfigure}{t}
\centering 
\includegraphics[scale=0.4]{figs/triangle.ps}
\caption{The (standard) unitarity triangle.\label{fig:ut}}
\end{nfigure}
The two non-trivial sides of the triangle 
are 
\begin{eqnarray}
  R_u &\equiv& \sqrt{\ov \rho^2 + \ov \eta^2 }, \qquad\qquad 
  R_t \;\equiv \; \sqrt{(1-\ov \rho)^2 + \ov \eta^2 }
\label{defsd} .
\end{eqnarray}
$C\!P$-violating quantities are associated with the triangle's three angles 
\begin{equation}
    \alpha = \arg\left[ - \frac{V_{td}V_{tb}^*}{V_{ud}V_{ub}^*}\right],
    \qquad
    \beta  = \arg\left[ - \frac{V_{cd}V_{cb}^*}{V_{td}V_{tb}^*}\right],
    \qquad
    \gamma = \arg\left[ - \frac{V_{ud}V_{ub}^*}{V_{cd}V_{cb}^*}\right].
    \label{eq:beta}
\end{equation}
The angle $\gamma$ coincides with $\delta_{13}$ of \eq{VPDG} at the
sub-permille level.  With \eqsto{defre}{eq:beta} one obtains
\begin{eqnarray}
 \ov{\rho} +  i \ov{\eta} &=&
              R_u  e^{i \gamma},\qquad\qquad 
1-\ov{\rho} - i \ov{\eta} \; = \; R_t  e^{-i \beta}
\label{defrb} .
\end{eqnarray}
The unitarity relation of \eq{sut} now simply reads 
\begin{eqnarray}
R_u  e^{i \gamma} +R_t  e^{-i \beta} &=& 1 \label{utrt}
\end{eqnarray}
Taking real and imaginary parts of \eq{utrt} reproduces formulae which
you know from high-school geometry, allowing us to express any two of
the four quantities $R_u,R_t,\gamma,\beta$ in terms of the remaining two
ones. By multiplying \eq{utrt} with either $\exp(-i\gamma)$ or
$\exp(i\beta)$ one finds analogous relations involving
$\alpha=\pi-\beta-\gamma$. 

Sometimes one needs to refine the Wolfenstein approximation to higher
orders in $\lambda$. It is prudent to define \cite{blo}
\begin{eqnarray}
\lambda \equiv s_{12}, &&\qquad\qquad   A \lambda^2 \equiv s_{23}
\label{prudent}
\end{eqnarray}
to all orders in $\lambda$ and to expand all CKM elements in terms of
$\lambda$, $A$, $\ov\rho$ and $\ov\eta$ to the desired order in
$\lambda$. Then, for example:
\begin{eqnarray}
V_{ub} &=&  A\lambda^3(\ov \rho-i \ov\eta) 
          \lt( 1+ \frac{\lambda^2}{2} + {\cal O} (\lambda^4) \rt). 
\label{vub}
\end{eqnarray}
The phase 
\begin{equation}
    \beta_s  = \arg\left[ - \frac{V_{ts}V_{tb}^*}{V_{cs}V_{cb}^*}\right]
        = \lambda^2 \ov \eta + O(\lambda^4)
    \label{betasdef}
\end{equation}
plays an important role in \bbms; $\beta_s$ is small, of order $0.02$
(equal to 1 degree). In the phase convention of \eq{VPDG} the phase of
$V_{cs}V_{cb}^*$ is ${\cal O} (\lambda^6)$ and
\begin{equation}
   \arg (-V_{ts}) = \beta_s ( 1+O(\lambda^2)) . \label{argvts}  
\end{equation} 
Organising the phases in powers of $\lambda$, we find all CKM elements real to
order $\lambda^2$ except for $V_{ub}$, $V_{td}$ and $V_{ts}$. Going to higher
orders one encounters $ \arg (-V_{cd})\simeq A^2 \ov\eta \lambda^4$ and $ \arg
(V_{cs})\simeq -A^2 \ov \eta \lambda^6$.

\subsection{Effective Hamiltonians}\label{sect:heff}
We now address the strong interaction, which is the main obstacle on our
way from quark diagrams to mesonic amplitudes like $M_{12}$ and $A(M\to
f)$.  In Sect.~\ref{sec:mqb} we have seen that weak processes of mesons
are multi-scale processes. For instance, \bbm\ involves three largely
separated scales, since $m_t \sim M_W \gg m_b \gg \lqcd$. These scales
must be disentangled to separate the short-distance QCD, which is
described by the exchange of quarks and gluons, from the long-distance
hadronic physics, whose characteristic property is the confinement of
quarks into hadrons. The key tool to separate the physics associated
with the scale $m_{\rm heavy}$ from the dynamics associated with $m_{\rm
  light}\ll m_{\rm heavy}$ is the construction of an \emph{effective field
  theory}. The corresponding \emph{effective Hamiltonian} $H^{\rm eff} $
is designed to reproduce the S-matrix elements of the Standard Model 
up to corrections of order $(m_{\rm light}/m_{\rm heavy})^n$ where $n$
is a positive integer:
\bea%
 \bra{f} \mathbf{T} e^{-i \int d^4 x H_{\rm int}^{\rm SM}(x)}
   \ket{i} &=&  \bra{f} \mathbf{T} e^{-i \int d^4 x H^{\rm eff}(x)}
   \ket{i} \lt[1 + 
   {\cal O} \lt( \frac{m_{\rm light}}{m_{\rm heavy}} \rt)^n \, \rt] 
  \label{eft} 
\eea%
I exemplify the method with an effective Hamiltonian which reproduces
the amplitude for \bbm\ up to corrections of order $m_b^2/M_W^2$. That
is, we employ \eq{eft} for the case $i=\Bbar$ and $f=B$ (where $B=B_d$
or $B_s$), $m_{\rm light}=m_b$ and $m_{\rm heavy}=M_W \sim m_t$. The
corresponding effective Hamiltonian reads%
\bea%
H^{\rm eff} &=& H^{\rm QCD (f=5)} + H^{\rm QED (f=5)} + H^{|\Delta
  B|=2} \label{hf5} .%
\eea%
Here the first two terms are the usual QCD and QED interaction
Hamiltonians with 5 ``active flavours'', meaning that they do not
involve the top quark. The last term describes the weak interaction.
Adapted to the process under study, $H^{|\Delta B|=2}$ only encodes the
physics related to \bbm, but does not describe other weak processes such
as meson decays. It is called $H^{|\Delta B|=2}$,
because it describes physical processes in which the bottom quantum
number $B$ changes by two units.  $H^{|\Delta B|=2}$ does not contain 
W-boson, Z-boson or top-quark fields, instead the $\Delta B=2$ transition 
of the box diagram in \fig{fig:boxes} is mediated by an effective
four-quark coupling: 
\begin{eqnarray}
  Q & =& 
    \ov{q}_L \gamma_{\nu} b_L \, \ov{q}_L \gamma^{\nu} b_L 
\qquad\qquad \mbox{with $q=d$ or $s$} .\label{defQ}
\end{eqnarray}    
For historical reasons $Q$ is called a \emph{four-quark operator}, but
it is nothing but a point-like coupling of four quark fields as shown in
\fig{fig:q}.
\begin{nfigure}{t}
\centering 
\includegraphics[scale=0.6]{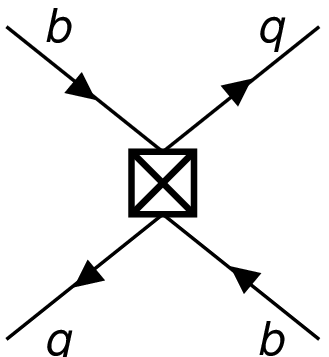}
\caption{The four-quark operator $Q$ for \bbmq\ with $q=d$ or $s$.
\label{fig:q}}
\end{nfigure}
We have 
\beq%
H^{|\Delta B|=2} = \frac{G_F^2}{4 \pi^2}\, ( V_{tb} V_{tq}^* )^2 \,
    C^{|\Delta B|=2}( m_t, M_W, \mu )\, Q (\mu) + h.c.
\label{ch1:h2}
\eeq%
where the lengthy expression multiplying $Q$ is just the effective
coupling constant multiplying the four-quark interaction of \fig{fig:q}.
This coupling constant is split into several factors, the first of which
contains the Fermi constant $G_F$. The second factor summarises the CKM
elements of the box diagram and the third factor $C^{|\Delta B|=2}( m_t,
M_W, \mu )$ is the \emph{Wilson coefficient}, which contains the
information on the heavy mass scales $M_W$ and $m_t$. Finally $\mu$ is
the renormalisation scale, familiar from QCD. Just as any other coupling also
$Q$ must be renormalised. The renormalised operator $Q$ depends on $\mu$
through the renormalisation constant $Z_Q(\mu)$ via $Q=Z_Q Q^{\rm bare}$
and (in a mass-independent scheme like $\ov{\rm MS}$) the latter
dependence is only implicit through $g(\mu)$, where $g$ is the QCD
coupling constant.\footnote{The analogy with the renormalisation of the
  QCD coupling constant is more obvious if one reads the product $C Z_Q
  Q^{\rm bare}$ in a different way: By assigning $Z_Q$ to $C$ rather
  than $Q$ one may view $C$ as a renormalised coupling constant. The
  notion of a ``renormalised'' operator instead of a ''renormalised
  Wilson coefficient'' has historical reasons. }  With the decomposition
in \eq{ch1:h2} $C^{|\Delta B|=2}$ has dimension two and is real.

$C^{|\Delta B|=2}$ is calculated from the defining property of $H^{\rm
  eff}$ in \eq{eft}: We compute the $\Delta B=2$ process both in the
Standard Model and with the interactions of $H^{\rm eff}$ and adjust
$C^{|\Delta B|=2}$ such that the two results are the same, up to
corrections of order $m_b^2/M_W^2$. Obviously we cannot do this with
mesons as external states $i$ and $f$. But a crucial property of $H^{\rm
  eff}$ is the independence of the Wilson coefficient on the external
states. We can compute it for an arbitrary momentum configuration for
the external quarks as long as the external momenta are of the order of
$m_{\rm light}$. That is, we do not need to know the complicated
momentum configuration of quarks bound in a meson state. Further all QCD
effects in $C^{|\Delta B|=2}$ are purely perturbative: \bea%
C^{|\Delta B|=2} &=& C^{|\Delta B|=2,(0)} + \frac{\alpha_s(\mu)}{4\pi}
C^{|\Delta B|=2,(1)} + \ldots
\label{cpert} 
\eea%
We can understand why and how this works if we expand the result of the
box diagram of \fig{fig:boxes} in terms of the momenta of the external
quarks, which are at most of order $m_b$. The leading term consists of
the result of a loop integral with external momenta set to zero and the
spinors of the external quark states. Now the ``effective theory side''
of \eq{eft} involves the tree-level diagram corresponding to%
\bea%
\bra{f} \mathbf{T} e^{-i \int d^4 x H^{\rm eff}(x)} \ket{i}^{(0)} 
  &\simeq & -i \int d^4 x
  \bra{f} H^{\rm eff}(x) \ket{i}^{(0)} 
  \; =\; -i \int d^4 x \bra{f} H^{|\Delta B|=2} (x) \ket{i}^{(0)} 
  \nn 
  & =& -i (2\pi)^4 \delta^{(4)}(p_f - p_i) \; \frac{G_F^2}{4
  \pi^2}\, ( V_{tb} V_{tq}^* )^2 \, C^{|\Delta B|=2,(0)} \, 
 \bra{f} Q \ket{i}^{(0)}  \no %
\eea%
where $\ket i=\ket{p_b,s_b; p_{\ov q}, s_{\ov q }}$ and $\ket f=
\ket{p_q,s_q; p_{\ov b}, s_{\ov b} } $ are the external states
characterised by the momenta and spins of the quarks. The superscript 
``$(0)$'' indicates the lowest order of QCD everywhere.  
Since $\bra{f} Q
\ket{i}$ reproduces the spinor structure (``Dirac algebra'') of the box
diagram, the coefficient $C^{|\Delta B|=2,(0)}$ inferred from this
\emph{matching calculation}\ is solely determined in terms of the loop
integral and therefore only depends on $M_W$ and $m_t$. The matching
calculation becomes less trivial when we go to the \emph{next-to-leading
  order (NLO)} of QCD. Now $H^{\rm QCD}$ enters the matching calculation
and we must dress both the box diagram and the effective diagram in
\fig{fig:q} with gluons in all possible ways. Denoting the SM amplitude
by%
\bea%
{\cal M} &=& {\cal M}^{(0)} + \frac{\alpha_s}{4\pi} {\cal M}^{(1)} +
\ldots, \label{sma} \eea%
our NLO matching calculation amounts to the determination of $C^{|\Delta
  B|=2,(1)}$ from%
\bea%
- {\cal M}^{(0)} - \frac{\alpha_s}{4\pi} {\cal M}^{(1)} &=& 
\frac{G_F^2}{4
  \pi^2}\, ( V_{tb} V_{tq}^* )^2 \, 
\lt[
C^{|\Delta B|=2,(0)} + \frac{\alpha_s }{4\pi} C^{|\Delta B|=2,(1)} \rt]
 \nn 
&& \qquad\qquad\cdot 
 \lt[ \langle Q \rangle^{(0)} + \frac{\alpha_s}{4\pi} \langle Q
\rangle^{(1)} \rt] \,
\lt[ 1+ {\cal O}\lt( \frac{m_b^2}{M_W^2} \rt) \rt] \; +\; {\cal O} \lt(
\alpha_s^2 \rt) \quad \label{manlo} %
\eea%
On the RHS the external states are suppressed for simplicity of
notation.  The QCD corrections to the box diagram in ${\cal M}^{(1)}$
not only depend on the light scales, i.e.\ external momenta and light
quark masses, they also suffer from infrared (IR) divergences. These
divergences signal the breakdown of QCD perturbation theory at low
energies. However, the gluonic corrections to \fig{fig:q}, which are
comprised in $\langle Q \rangle^{(1)}$, exactly reproduce the infrared
structure of the SM diagrams: They involve the same IR divergences and
have the same dependence on the light mass scales. Collecting the 
${\cal O}(\alpha_s)$ terms from \eq{manlo},%
\bea%
- {\cal M}^{(1)} &=& 
\frac{G_F^2}{4 \pi^2}\, ( V_{tb} V_{tq}^* )^2 \, 
\lt[ C^{|\Delta B|=2,(0)} \langle Q \rangle^{(1)}  + 
     C^{|\Delta B|=2,(1)}  \langle Q \rangle^{(0)} \rt],
\label{manlocoll} %
\eea%
one finds identical IR structures on the LHS and in the first term in the
square brackets, while $ C^{|\Delta B|=2,(1)}$  only contains heavy masses and
no IR divergences.  In conclusion, the IR structure of the SM amplitude
properly factorises with an ``infrared-safe'' $ C^{|\Delta B|=2}$. This
success can be understood by separately discussing the regions of small and
large loop momentum passing through a gluon line in the diagrams of ${\cal
  M}^{(1)}$.  The infrared-sensitive diagrams are identified as those in which
the gluon connects two external quark lines. (The other diagrams are
infrared-finite and one can set the light mass parameters to zero.)  If the
loop momentum traversing the gluon line is small, we can neglect it in the
heavy top and W propagators. Therefore the loop integration factorises into
two one-loop integrations and the second loop integral involving the heavy
particles simply reproduces the one-loop result contained in $ C^{|\Delta
  B|=2,(0)}$. The gluon-loop integration ---still over soft momenta only--- is
equal to the one in the corresponding diagram in $\langle Q\rangle^{(1)}$,
where the gluon connects the same quark lines. Therefore the region of
integration with a soft gluon factorises with the leading-order coefficient
$C^{|\Delta B|=2,(0)}$ in \eq{manlo}.  The region of the momentum integration
with a hard gluon momentum does not factorise in this way and contributes to
$C^{|\Delta B|=2,(1)}$.  However, the region of large gluon loop momentum is
not infrared-sensitive and we can neglect the light momenta and masses.
Therefore $C^{|\Delta B|=2,(1)} $ does not depend on the light mass scales.
Conversely, $\langle Q \rangle$ contains only small scales of order $m_{\rm
  light}$ and encodes the full infrared structure of ${\cal M}$. Therefore our
quark-level calculation is meaningful for $C^{|\Delta B|=2} $, but not for
$\langle Q \rangle$. In order to make a theoretical prediction for the \bbm\ 
amplitude, we must compute $\bra{B} Q \ket{\,\ov{\!B}}$ with nonperturbative
methods.  The factorisation of ${\cal M}$ into short-distance coefficients and
long-distance operator matrix elements is also called \emph{operator product
  expansion}.

Here I only derive the result for the leading-order (LO) Wilson 
coefficient $C^{|\Delta  B|=2,(0)} $. In a first step let us decompose 
${\cal M}^{(0)}$ as%
\beq%
   {\cal M}^{(0)} = \sum_{j,k=u,c,t} 
   V_{jb}^*V_{jq}\, V_{kb}^*V_{kq}\, {\cal M }^{(0)}_{jk} 
    \langle Q \rangle ^{(0)}, 
   \qquad\qquad \mbox{$q=d$ or $s$,}  \label{loij}
\eeq%
where ${\cal M }^{(0)}_{jk} \langle Q \rangle ^{(0)} $ 
is the result of the box diagram containing
internal quark flavours $(j,k)$ with the CKM elements factored out.
We then write
\bea%
   {\cal M }^{(0)}_{jk} &=& - \frac{G_F^2}{4 \pi^2}\, M_W^2\,
    \widetilde S  (x_j, x_k ) \label{stil} 
\eea%
with $x_j=m_j^2/M_W^2$. The function $ \widetilde S (x_j, x_k )$ is
symmetric, $ \widetilde S (x_j, x_k )= \widetilde S (x_k, x_j )$.   
In the next step we use CKM unitarity to 
eliminate $V_{ub}^*V_{uq}= - V_{tb}^*V_{tq} - V_{cb}^*V_{cq}$ from 
\eq{loij}: 
\beq%
  - {\cal M}^{(0)} =  \frac{G_F^2}{4 \pi^2}\, M_W^2\, 
   \lt[  
     \lt( V_{tb}^*V_{tq} \rt)^2 S(x_t) \,+\, 
      2 V_{tb}^*V_{tq}\, V_{cb}^*V_{cq} S(x_c,x_t) \, + \,    
     \lt( V_{cb}^*V_{cq} \rt)^2 S(x_c) \rt] \, 
    \langle Q \rangle ^{(0)}. \label{sij}
\eeq%
$S$ and $\widetilde S$ are related as%
\bea%
  S( x_j, x_k ) &=& 
             \widetilde  S (x_j,x_k ) - \widetilde S (x_j,0 ) 
              - \widetilde S (0,x_k ) + \widetilde S (0, 0),
 \qquad \mbox{for $j,k=c,t$},\nn
   S(x) &\equiv &S(x,x) , \label{inali}%
\eea%
where I have set the up-quark mass to zero. In \eq{sij} the last two
terms are tiny, because $x_c \sim 10^{-4}$ and%
\beq%
  S(x_c)= {\cal O} (x_c), \qquad\qquad  
  S(x_c,x_t)= {\cal O} (x_c \ln x_c). \label{ilexp}%
\eeq%
This consequence of CKM unitarity is called the
\emph{Glashow-Iliopoulos-Maiani (GIM)} suppression, related to the
vanishing of FCNCs in the limit of equal internal quark masses (here
$m_c$ and $m_u=0$). No GIM suppression occurs in top loops, because 
$x_t \sim 4 $. The dominant contribution to \eq{loij} involves%
\bea%
S(x_t) &=& x_t \left[ \frac{1}{4} + \frac{9}{4}
    \frac{1}{1-x_t} - \frac{3}{2} \frac{1}{(1-x_t)^2} \right]
    - \frac{3}{2} \left[ \frac{x_t}{1-x_t} \right]^3 \ln x_t 
 \;\approx \; 2.3  .
 \label{sxt} %
\eea%
The tiny charm contribution does not contribute to 
$C^{|\Delta  B|=2,(0)}$ at all; to accommodate for it we must refine our 
operator product expansion to include higher powers of $(m_{\rm
  light}/m_{\rm heavy})$ in \eq{eft}. We can read off $C^{|\Delta
  B|=2,(0)}$ from \eqsand{manlo}{sij}:%
\bea%
C^{|\Delta B|=2,(0)} ( m_t, M_W, \mu ) = M_W^2\,
    S\, ( x_t ) . \label{wcini}
\eea%
The functions $S(x)$ and $S(x_c,x_t)$ are called \emph{Inami-Lim}\ 
functions \cite{il}.

The factorisation in \eqsand{eft}{manlo} also solves another problem: %
No largely separated scales appear in $C^{|\Delta B|=2}( m_t, M_W, \mu
)$ provided that we take $\mu = {\cal O} (M_W,m_t)$, so that no large
logarithms can spoil the convergence of the perturbative series. While
no explicit $\mu$-dependence is present in our LO result in \eq{wcini},
there is an implicit $\mu$-dependence through $m_t(\mu)$, which is a
running quark mass (typically defined in the $\ov{\rm MS}$ scheme).
$C^{|\Delta B|=2,(1)}$ also contains an explicit $\ln (\mu/M_W)$ term.  
Two sources contribute to this term: First, there is already a 
$\ln (\mu/M_W)$ term in $ {\cal M}^{(1)}$, familiar to us from 
matrix elements with  $\ov{\rm MS}$-renormalised UV divergences. Second, 
$ {\cal M}^{(1)}$ contains the large logarithm $\ln(m_b/M_W)$ which is 
split between matrix elements and Wilson coefficients as 
\bea%
       \ln \frac{m_b}{M_W} &=&  
        \ln \frac{m_b}{\mu} + \ln \frac{\mu}{M_W} . \label{logs}
\eea%
This feature is transparent from \eq{manlocoll}.

The scale $\mu_{tW}= {\cal O} (M_W,m_t)$ at which we invoke \eq{manlo} to
find $ C^{|\Delta B|=2}$ is called the \emph{matching scale} and
$C^{|\Delta B|=2}( m_t, M_W, \mu_{tW} )$ has a good perturbative
behaviour. Similarly, no large logarithms occur in $\langle Q (\mu_b)
\rangle$, if we choose a scale $\mu_b\sim m_b$ in the matrix element.
Since the $\mu$-dependence in $H^{|\Delta B|=2}$ is spurious, we can
take any value of $\mu$ we want, but this value must be the same in
$C(\mu)$ and $\langle Q (\mu) \rangle$. That forces us to either relate
$C(\mu_{tW})$ to $C(\mu_b)$ or to express $\langle Q (\mu_b) \rangle $
in terms of $\langle Q (\mu_{tW}) \rangle $ in such a way that large
logarithms%
\beq%
\alpha_s^n \ln^n \frac{\mu_{tW}}{\mu_b} \eeq%
are summed to all orders $n=0,1,2\ldots$ in perturbation theory.  This
can be achieved by solving the \emph{renormalisation group (RG)
  equation} for either $C(\mu)$ or $\langle Q (\mu) \rangle$. All steps
of this procedure are analogous to the calculation of the running quark
mass, which can be found in any textbook on QCD. RG-improvement promotes
our LO result to a \emph{leading-log (LL)}\ quantity:%
\bea%
C^{|\Delta B|=2,(0)} ( m_t, M_W, \mu_b ) 
  &=& u^{(0)}(\mu_b, \mu_{tW}) C^{|\Delta B|=2,(0)} ( m_t, M_W, \mu_{tW} )     
      \label{rg1} \\
 \langle Q (\mu_{tW})\rangle &=&  u^{(0)}(\mu_b, \mu_{tW}) 
       \langle Q (\mu_b) \rangle \label{rg2} \\ 
 u^{(0)}(\mu_b, \mu_{tW}) &=& \lt(
   \frac{\alpha_s(\mu_{tW})}{\alpha_s(\mu_b)} \rt)^{
     \frac{\gamma_+^{(0)}}{2\beta_0^{(5)}}} 
  \qquad\qquad \mbox{with } \gamma_+^{(0)} = 4 
  .\label{rg3}   
\eea%
The evolution factor $ u^{(0)}(\mu_b, \mu_{tW}) $ depends on the
\emph{anomalous dimension}\ of $Q$, which equals $(\alpha_s/(4\pi))
\gamma_+^{(0)}$ to LL accuracy. $\beta_0^{(f)}=11-2f/3$ is the first term
  of the QCD $\beta$ function. One usually writes 
\bea%
C^{|\Delta B|=2} ( m_t, M_W, \mu_b )
 &=& \eta_B b_B(\mu_b) C^{|\Delta B|=2,(0)} ( m_t, M_W, \mu_{tW} )
\label{ceb}
\eea%
where all dependence on $\mu_b$ is absorbed into $b_B(\mu_b)$ and all 
heavy scales reside in $\eta_B$.  This
factorisation is possible to all orders in $\alpha_s$. It is trivially
verified in the LL approximation of \eq{rg3}, where simply $u^{(0)} (\mu_b,
\mu_{tW}) =\eta_B b_B(\mu_b)$.  In \eq{ceb} $m_t$ is understood as $m_t(m_t)$
(and not as $m_t(\mu_{tW})$). In this way $\eta_B$ is independent of $\mu_{tW}$
to the calculated order; the residual $\mu_{tW}$ dependence is already tiny in
the NLL result. $\eta_B$ mildly depends on $x_t=m_t^2/M_W^2$ and in
practice one can treat it as a constant number \cite{bjw}: %
\bea%
  \eta_B = 0.55,\qquad\qquad b_B(\mu_b=m_b=4.2\,\gev) = 1.5 . 
   \label{ebnum}%
\eea%
The dependences of $b_B$ on $\mu_b$ and the chosen renormalisation
scheme cancel in the product $ b_B (\mu_b) \langle Q (\mu_b) \rangle$.
The quoted number is for the $\ov{\rm MS}$--NDR scheme, where ``NDR''
refers to the treatment of the Dirac matrix $\gamma_5$. Details on this
topic can be found in \cite{bw}.  We see that the impact of
short-distance QCD corrections is moderate, since
$\eta_B\,b_B(\mu_b)=0.84$. The NLL calculation of Ref.~\cite{bjw} has
found only small two-loop corrections and the remaining uncertainty
affects $\eta_B$ only in the third digit behind the decimal point.
RG-improved perturbation theory works superbly!  Combining
Eqs.~(\ref{ch1:h2}), (\ref{wcini}) and (\ref{ceb}) we obtain our final
expression for the $|\Delta B|=2$ hamiltonian:
\begin{equation}
H^{|\Delta B|=2} \; = \; \frac{G_F^2}{4 \pi^2}\, M_W^2\, 
( V_{tb} V_{tq}^* )^2 \, \eta_B \, S (x_t) 
   b_B(\mu_b) Q(\mu_b) 
  \; + \; h.c. \label{desh2}
\end{equation}

Finally we cannot escape from quark confinement! Our hadronic matrix
element is conventionally parameterised as%
\bea%
\bra{B_q} Q(\mu_b) \ket{\,\ov{\!B}_q} & =& \frac{2}{3} M_{B_q}^2 \,
f_{B_q}^2 \, \frac{\widehat B_{B_q}}{ b_B(\mu_b)} \label{mel}%
\eea%
with the $B_q$ meson decay constant $f_{B_q}$ and the \emph{bag factor}\
$\widehat B_{B_q}$. The parameterisation in \eq{mel} is chosen in such a
way that $\widehat B_{B_q}/b_B(\mu_b)$ is close to one. It will be especially
useful once precise experimental data on $f_{B_d}\sim f_{B^+}$ from
leptonic $B^+$ decays will be available.  With the help of our effective
field theory we have beaten the problem of long-distance QCD in \bbm\
down to the calculation of a single number. Lattice gauge theory
computations cover the ranges \cite{latt}%
\bea%
f_{B_d} \sqrt{\widehat B_{B_d}} & =& (225 \pm 35)\, \mev, \qquad\qquad
f_{B_s} \sqrt{\widehat B_{B_s}} \; =\; (270 \pm 45)\, \mev .
\label{latt} %
\eea%
The quoted hadronic uncertainties are the main problem in the extraction
of $|V_{tb}V_{tq}|$ from the measured $\dm_{B_q}$. $\widehat B_{B_d}$
could differ from $\widehat B_{B_s}$, but no computation has established
any significant difference by now.

Putting \eqsand{desh2}{mel} together we find the desired 
element of the \bb\ mass matrix: %
\bea%
M_{12} & = & 
   \frac{\bra{B_q} H^{|\Delta B|=2}  \ket{\,\ov{\!B}{}_q} }{2 M_{B_q}} \nn
& =&   \frac{G_F^2}{12 \pi^2}\, \eta_B\, M_{B_q} \,
     \widehat{B}_{B_q} f_{B_q}^2 \,
    M_W^2\, S \bigg( \frac{m_t^2}{M_W^2} \bigg)
    \left( V_{tb} V_{tq}^* \right)^2 .
    \label{m12b}%
\eea%
We remark that there is no contribution of $H^{|\Delta B|=2}$ to
$\Gamma_{12}$, because $\bra{B_q} H^{|\Delta B|=2} \ket{\,\ov{\!B}{}_q}$
has no absorptive part. By inspecting \eq{absdisp} we can verify that
the dispersive or absorptive part of some amplitude can be calculated by
replacing the loop integrals by their real or imaginary parts,
respectively, while keeping all complex CKM elements. But only diagrams
with light internal quarks involve loop integrals with a non-zero
imaginary part.  Hence we must extend our effective-Hamiltonian
formalism to include the effects of light internal quarks in the box
diagrams, if we want to predict $\dg_{B_q}$. Contracting the heavy
W-boson lines in the diagrams of \fig{fig:boxes} to a point does not
correspond to a contribution from $H^{|\Delta B|=2}$ in the effective
theory. Instead this is a second-order effect involving some effective
$|\Delta B|=1$-Hamiltonian $H^{|\Delta B|=1}$, which we must add to
$H^{\rm eff}$ in \eq{hf5}. The relevant piece from the RHS of \eq{eft}
is%
\bea%
- \frac{1}{2} \int d^4 x d^4 y \; \bra{B} {\bf T} H^{|\Delta B|=1} (x)
H^{|\Delta B|=1} (y) \ket{\,\ov{\!B}} . \label{bilo} \eea%
The LO contribution to this bilocal matrix element is depicted in
\fig{fig:dega} for the case of \bbms.
\begin{nfigure}{t}
\centering 
\includegraphics*[scale=0.7]{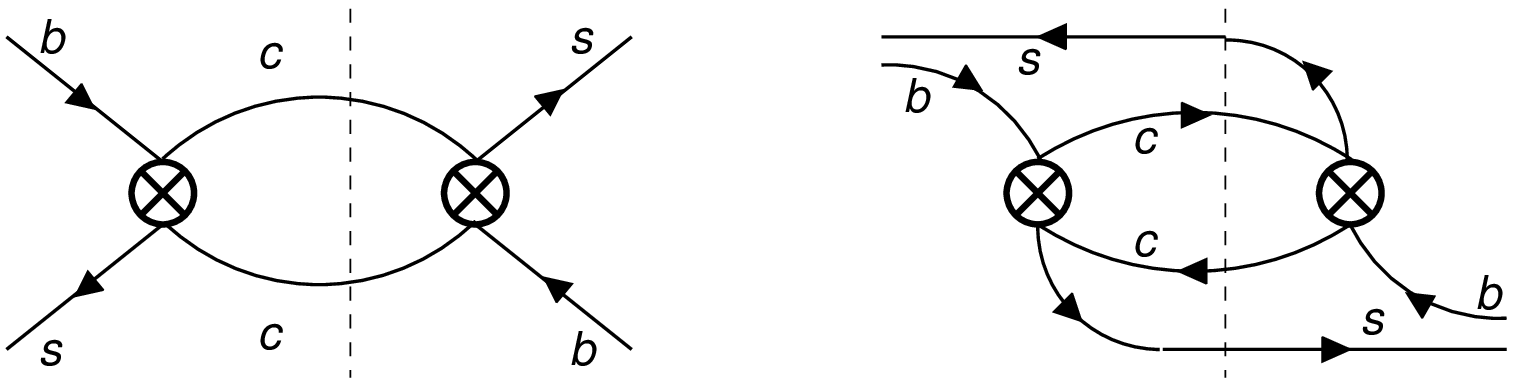}
\caption{Second-order contribution of $H^{|\Delta B|=1}$ 
   to \bbms. The diagrams constitute the dominant contribution to 
  $\dg_{B_s}$. }\label{fig:dega}
\end{nfigure}
The contribution from \eq{bilo} to \bbm\ is much smaller than the one from 
$H^{|\Delta B|=2}$, which is enhanced due to the heavy top mass entering 
\eq{sxt}. Therefore we can neglect the bilocal contribution in 
$M_{12}$ and only need to consider it for $\Gamma_{12}$. From this
observation we also conclude that $|\Gamma_{12}|\ll|M_{12}|$ leading to 
$|\dg|\ll \dm$, which we already exploited in \eqsto{agmb}{qpb}.   

\boldmath   
\subsection{SM predictions of \dm, \dg\ and $a_{\rm fs}$}
\unboldmath   
In Sec.~\ref{sect:heff} we have collected all ingredients of the SM
calculation of $\dm=2|M_{12}|$ for the $B_d$ and $B_s$ systems. Looking
at \eq{wolf} we realise that $|V_{tb}|$ is well-known and $|V_{ts}|$ 
is essentially fixed by the well-measured $|V_{cb}|$. From 
\eqsand{m12b}{latt} we find the SM prediction
\beq%
 \dm_{B_s} = \left( 12.5 \pm 4.3 \right) \, \mbox{meV} 
           = \left( 19.0 \pm 6.6 \right) \, \mbox{ps}^{-1} .
 \label{smdms}
\eeq
The first unit is milli-electronvolt, a unit which we do not encounter
often in high-energy physics. By dividing with $\hbar$ one finds the
second expression in terms of inverse picoseconds, which is more useful
since $\dm$ is measured from the oscillation frequency in \eq{defa0}. 
\eq{smdms} is in good agreement with the Tevatron measurement of 
\cite{cdf,d0mix}
\beq%
 \dm_{B_s}^{\rm exp} 
  = \left( 17.77 \pm 0.10_{\rm (stat)} \pm 0.07_{\rm (syst)} 
         \right) \, \mbox{ps}^{-1}
 . \label{smdmsexp}
\eeq%
The corresponding quantity for \bbmd\ is well-measured by several
experiments with \cite{pdg}
\beq%
 \dm_{B_d}^{\rm exp} = \left(333.7  \pm 3.3 \right) \, \mbox{$\mu$eV}
 = \left( 0.507 \pm 0.005 \right) \, \mbox{ps}^{-1}
 . \label{smdmdexp} 
\eeq%
We can use $ \dm_{B_d}$ to determine $|V_{td}|$. From \eq{m12b} 
we infer  
\beq%
 \dm_{B_d}  = 
(0.52\pm 0.02) \, \mbox{ps}^{-1}
          {\lt( \frac{|V_{td}|}{0.0082}\rt)^2 } \;
      \lt( \frac{f_{B_d} \sqrt{\widehat B_{B_d}}}{225 \, \mbox{MeV}} \rt)^2
\label{smdmd} .%
\eeq%
The 16\% error of the lattice value in  \eq{latt} dominates the 
uncertainty on the extracted $|V_{td}|$. The 
all-order Wolfenstein parameterisation defined by 
\eqsand{defre}{prudent} implies  
\beq%
  |V_{td}| = A \lambda^3 R_t + {\cal O} (\lambda^5).
\eeq%
Since $A\lambda^2\simeq |V_{cb}|$ is well-known, $\dm_{B_d}$
essentially determines $R_t$, i.e.\ one side of the unitarity triangle.
Even better, we can use the ratio $\dm_{B_d}/\dm_{B_s}$ for the same
purpose: If one forms the ratio of the hadronic quantities in \eq{latt},
many uncertainties drop out:
\begin{eqnarray}
 \xi & =& 
   \frac{f_{B_s} \sqrt{\widehat{B}_{B_s}}}{
              f_{B_d} \sqrt{\widehat{B}_{B_d}}}
      \;=\; 1.20 \pm 0.06 .\label{xi}
\end{eqnarray}
In the limit of exact flavour-SU(3) symmetry (corresponding to $m_u=m_d=m_s$)
one has $\xi=1$ which reduces the calculational task to compute the deviation
of $\xi$ from 1. The somewhat large error in \eq{xi} reflects the ongoing
discussion on potentially large chiral logarithms \cite{kr} which may increase
$\xi$ significantly. This problem occurs, because lattice simulations use
values for the pion mass which are larger than the physical value. The
extrapolation to $m_\pi\simeq 140\,\mev$ with the help of chiral perturbation
theory introduces this source of error.  Sum-rule calculations of $\xi$ (or
rather $f_{B_s}/f_{B_d}$) which automatically include these logarithms,
however, give values at the lower end of the range in \eq{xi} \cite{jl}.
Further all short-distance QCD drops out from the ratio $\dm_{B_d}/\dm_{B_s}$,
so that one simply has
\begin{eqnarray}
\lt| \frac{V_{td}}{V_{ts}} \rt| &=&
   \sqrt{\frac{\dm_{B_d}}{\dm_{B_s}}} \,
   \sqrt{\frac{M_{B_s}}{M_{B_d}}} \, \xi . \label{vdmds}
\end{eqnarray}
The Wolfenstein expansion leads to 
\begin{eqnarray}
\lt| \frac{V_{td}}{V_{ts}} \rt| &=& 
  R_t \lambda \lt[ 1 + \lambda^2 \lt( \frac12 -\ov \rho \rt)  
                     + {\cal O} (\lambda^4) \rt].  
 \label{vtdtsexp}
\end{eqnarray}
Combining \eqsand{vdmds}{vtdtsexp} (and using $M_{B_s}/M_{B_d}= 1.017$) 
we easily derive a home-use formula for $R_t$:
\begin{eqnarray}
R_t &=& 0.887 \, \frac{\dm_{B_d}}{0.507 \,\mbox{ps}^{-1}}  
              \, \frac{17.77 \,\mbox{ps}^{-1}}{\dm_{B_s}}
              \, \frac{\xi}{1.2} \, \frac{\lambda}{0.2246} 
              \, \lt[\, 1 + 0.05 \, \ov \rho \, \rt] 
  \label{home}
\end{eqnarray}
Neither $\ov \rho \approx 0.2$ nor the 1\% error on $\lambda\simeq 0.2246$ 
have an impact on the error of $R_t$. Using the numerical input 
from \eqsto{smdmsexp}{smdmdexp} and \eq{xi}  we find  
\begin{eqnarray}
   R_t \; =\; 0.90 \pm 0.04 \label{rtres} 
\end{eqnarray}
and the uncertainty is essentially solely from $\xi$ in \eq{xi}. 

Next we discuss \dg\ and the quantity $a_{\rm fs}$ in \eq{afst}, which
governs $C\!P$ violation in mixing. In order to find these quantities we
need to calculate $\Gamma_{12}$. This involves the diagrams of
\fig{fig:dega} and brings in a new feature, power corrections of order
$\lqcd/m_b$ \cite{bbd1}.  NLL QCD corrections to $\Gamma_{12}$ in the B
system have been calculated in Ref.~\cite{bbgln1,bbln,rome03}. In the SM
the $C\!P$ phase $\phi$ of \eq{defphi} is so small that one can set
$\cos\phi$ to 1 in \eq{mgsol:b}.  If we normalise $\dg$ to $\dm$ we can
eliminate the bulk of the hadronic uncertainties.  Updated values,
obtained by using an improved operator basis, are \cite{ln}
\begin{eqnarray}
\dg_{B_s} & = &
\left( \frac{\dg_{B_s}}{\Delta M_{B_s}} \right)^{\rm th}
\, \Delta M_{B_s}^{\rm exp} \;=\; 0.088 \pm 0.017
\, \mbox{ps}^{-1} ,\label{dgsnum} \\
\dg_{B_d} &=& \lt( \frac{\dg_{B_d}}{\dm_{B_d}} \right)^{\rm th}
\, \dm_{B_d}^{\rm exp} \; =\;
     \lt( 26.7 \epm {5.8}{6.5}\rt) \cdot 10^{-4} \, \mbox{ps}^{-1} . 
 \label{dgdnum}
\end{eqnarray}
The width difference in the $B_s$ system amounts to $12.7 \pm 2.4\%$ of the
average width $\Gamma_{B_s}\simeq \Gamma_{B_d}$ \cite{ln} and is in the reach
of present experiments.  Needless to say that there are no useful data on
$\dg_{B_d}$.  The predictions for the $C\!P$ asymmetries in flavour-specific
decays of \eq{afst} are calculated from \eq{agmb} and read
\cite{bbln,rome03,ln}
\begin{eqnarray}
a_{\rm fs}^s & = & \left( 2.06 \pm 0.57 \right) \cdot 10^{-5} 
\label{afssnum}\\
a_{\rm fs}^d & = & \lt(  -4.8\epm{1.0}{1.2} \rt) \,
                \cdot 10^{-4}. \label{afsdnum}
\end{eqnarray}
Also the current data for these  $C\!P$ asymmetries  are not useful for 
CKM metrology. A future measurement of $a_{\rm fs}^{d\, \rm exp}$ will add an 
interesting new constraint to the $(\ov \rho,\ov \eta)$ plane \cite{bbln}:
\begin{eqnarray}
  (\ov \eta - R_{\rm fs})^2 + (1 - \ov \rho)^2 \; =\; R_{\rm fs}^2
  && \qquad\qquad \mbox{with }\qquad 
      R_{\rm fs} = - \frac{a_{\rm fs}^{d\, \rm exp}}{ 
                           \lt(10.1\epm{1.8}{1.7}\rt) \cdot 10^{-4}} 
  . \label{circ} 
\end{eqnarray}
The theory prediction of Refs.~\cite{bbln,rome03} enters the denominator of 
$R_{\rm fs}$, the quoted value is consistent with \eq{afsdnum} 
and stems from the update in Ref.~\cite{ln}. 
\eq{circ} defines a circle with radius $R_{\rm fs}$ centred around 
$(\ov \rho,\ov \eta)=(1,R_{\rm fs})$. Therefore the circle 
touches the $\ov \rho$ axis at the point $(1,0)$, see \fig{fig:circ}.    
\begin{nfigure}{t}
\centering 
\includegraphics*[scale=0.7]{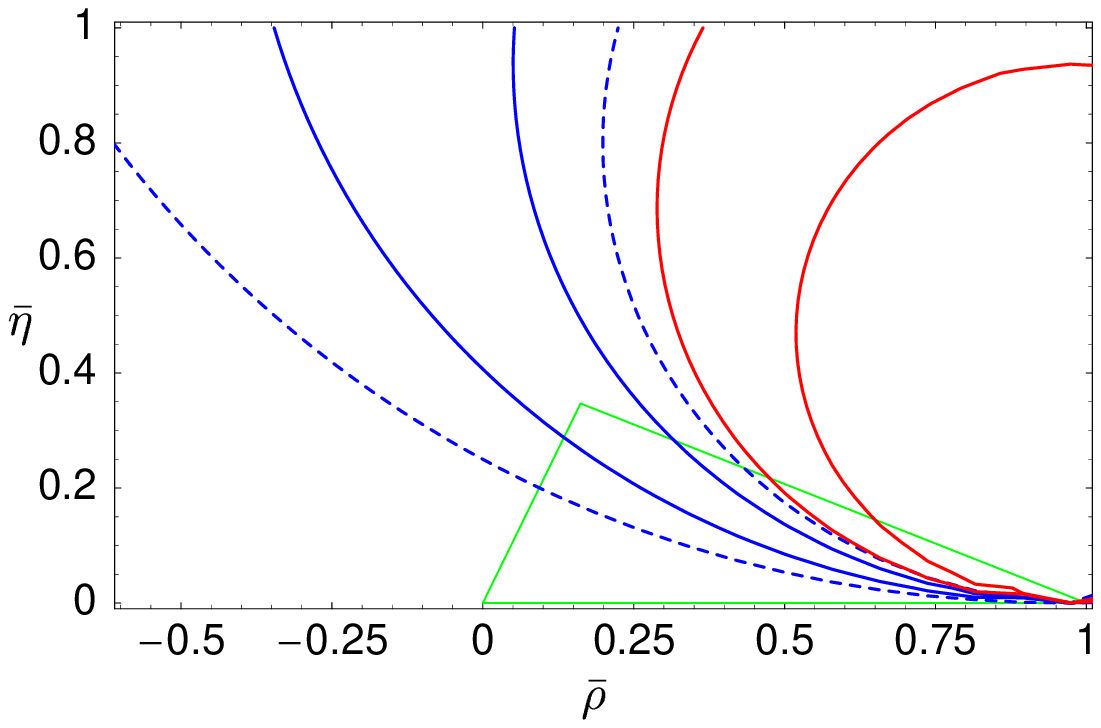}
\caption{Impact of $a_{\rm fs}^d$ on the $(\ov \rho,\ov \eta)$ plane:
  The solid blue curves limit the allowed range (defined by the error in
  \eq{circ}) for a hypothetical measurement of $a_{\rm fs}^{d\, \rm
    exp}=-5\cdot 10^{-4}$. The solid red curves are for $a_{\rm fs}^{d\, \rm
    exp}=- 10^{-3}$ instead.  For further information see Ref.~\cite{bbln},
  from which the figure is taken.}\label{fig:circ}
\end{nfigure}

We have seen that the three quantities related to \bbms\ discussed in
Eqs.~(\ref{smdms}), (\ref{dgsnum}) and (\ref{afssnum}) have little dependence
on $\ov \rho$ and $\ov \eta$. Only $\dm_{B_s}$ has an impact on CKM metrology,
through \eq{home}. The small sensitivity to $\ov \rho$ and $\ov \eta$ becomes
a virtue in searches for new physics, where \bbms\ plays an important role.

Next we discuss \kkm: The calculation of $M_{12}$ now forces us to compute box
diagrams of \fig{fig:boxes} with all possible quark flavours $u,c,t$, because
the top contribution involving $S(x_t)$ is suppressed by the small CKM factor
$ ( V_{ts}^*V_{td} )^2 \simeq A^4 \lambda^{10} (1-\ov\rho +i \ov \eta)^2$. The
charm and up contributions, however, are proportional to only two powers of
$\lambda$. Therefore we cannot neglect these contributions despite of the
smallness of $S(x_c)$ and $S(x_c,x_t)$ (discussed around \eq{ilexp}). Their
calculation proceeds in two major steps: First, the top quark and W-boson are
integrated out. In the resulting effective theory the $\Delta S=2$ transitions
receive second-order contributions from a $|\Delta S|=1$-Hamiltonian
$H^{|\Delta S|=1}$.  We have already seen this in our discussion of $\Delta
B=2$ transitions, the corresponding expression for \kkm\ is obtained by
replacing $H^{|\Delta B|=1}$ with $H^{|\Delta S|=1}$ in \eq{bilo} (and is
described by the analogous diagrams of \fig{fig:dega}). In addition to this
bilocal contribution, the term with $S(x_c,x_t)$ also involves a $|\Delta
S|=2$-Hamiltonian $H^{|\Delta S|=2}$ which mediates \kkm\ via a local
four-quark operator, just as in the case of \bbm. The $|\Delta S|=1$ and
$|\Delta S|=2$ Wilson coefficients of this effective field theory are evolved
down to the scale $\mu_{bc}={\cal O}(m_c)$ at which the second step of the
calculation is performed: Now the bottom and charm quarks are integrated out
and the effective field theory set up in the first step is matched to another
effective field theory. The new theory treats $m_b$ and $m_c$ as heavy scales,
so that all box diagrams involving at least one charm quark are effectively
contracted to a point. All information on $m_c$ (and $m_b$ which plays a minor
role) resides in the Wilson coefficient of the local $\Delta S=2$ operator
\begin{eqnarray}
  Q & =& 
    \ov{d}_L \gamma_{\nu} s_L \, \ov{d}_L \gamma^{\nu} s_L .  
\label{defQk}
\end{eqnarray}    
The effective $|\Delta S|=2$ Hamiltonian can therefore be written in  a
similar way as the $|\Delta B|=2$ Hamiltonian of \eq{desh2}:
\begin{eqnarray}
H^{|\Delta S|=2} & = & \frac{G_F^2}{4 \pi^2}\, M_W^2 \, 
  \lt[ ( V_{ts} V_{td}^* )^2 \, \eta_{tt} \, S ( x_t ) \, + \,
      2 V_{ts} V_{td}^* V_{cs} V_{cd}^*  \, \eta_{ct} \, 
                                      S (x_c, x_t ) \rt. \nn 
&& \lt. \qquad \qquad \qquad \qquad \, + \,
       ( V_{cs} V_{cd}^* )^2 \, \eta_{cc} \, x_c  \rt] \,
   b_K(\mu_K) Q(\mu_K) 
  \; + \; h.c. \label{deshs2}
\end{eqnarray}
The NLL results for the short-distance QCD factors read
\begin{equation}
\eta_{tt} = 0.57, \qquad 
\eta_{ct} = 0.47 \pm 0.05, \qquad
\eta_{cc} = (1.44 \pm 0.35) \lt(\frac{1.3\,\gev}{m_c}\rt)^{1.1} 
\label{etas} . 
\end{equation}
The QCD coefficients in \eq{etas} were calculated to LL accuracy in
Ref.~\cite{vzns}.  
The NLL calculation of $\eta_{tt}$ \cite{bjw} is analogous to
that of $\eta_B$, with one new feature: When crossing the threshold $\mu_{bc}$
one must change the number of active flavours in the QCD $\beta$ function and
the NLL anomalous dimension $\gamma_+$ from $f=5$ to $f=3$. The NLL results
for $\eta_{ct}$ \cite{hn1} and $\eta_{cc}$ \cite{hn2} have a sizable
uncertainty, because they are sensitive to the low scale of $\mu_{bc}\sim m_c$
where $\alpha_s$ is large. $\eta_{cc}$ also exhibits a sizable dependence on
$\alpha_s(M_Z)$ and on $m_c=m_c(m_c)$, so that the central values quoted in
the literature vary over some range. The expression in \eq{etas} approximates
the dependence on $m_c$ and corresponds to $\alpha_s(M_Z)=0.119\pm 0.002$.
The scale $\mu_K$ must be chosen below $m_c$ and is typically taken around
$1\,\gev$, where perturbation theory is still applicable.  One finds
$b_K(\mu_K=1\,\gev)=1.24\pm 0.02$ and the error stems from the uncertainty in
$\alpha_s$.

In the discussion of $|\Delta S|=2$ transitions we must also address 
corrections of order $m_{\rm light}^2/m_{\rm heavy}^2$ which correspond to
subleading terms in the operator product expansion of \eq{eft}. While these
corrections are of order $\lqcd^2/m_t^2$ for the first term in 
$H^{|\Delta S|=2}$, they are of order $\lqcd^2/m_c^2$ in the case of the 
charm contributions involving $S(x_c,x_t)={\cal O} (x_c \ln x_c)$ and 
$S(x_c)\simeq x_c $ in \eq{deshs2}. The largest of these power corrections
involves two $|\Delta S|=1$ operators and corresponds to the box diagram 
in \fig{fig:boxes} with two internal up-quarks. To understand the power
counting, recall that the charm contribution in $H^{|\Delta S|=2}$ is
proportional to $M_W^2 x_c=m_c^2$, while the box with up-quarks involves 
no power of $m_c$, so that its size is characterised by the hadronic 
energy scale $\lqcd$. Including this bilocal contribution we write:%
\beq%
M_{12} = \frac{1}{2 m_K}\, \bra{K} H^{|\Delta S|=2} \ket{\,\ov{\!K}}
    - \mbox{Disp}\, \frac{i}{4 m_K} \int \! d^4 x \,
    \bra{K} H^{|\Delta S|=1} (x)\, H^{|\Delta S|=1} (0) \ket{\,\ov{\!K}} \,.
\label{m12k}
\eeq%
Here ``Disp'' denotes the dispersive part of the matrix element, which is 
introduced in \eq{absdisp} and is discussed after \eq{m12b}. 
The enhancement of the second term stems from the so-called \emph{$\Delta I=0$
  rule}\ which describes the non-perturbative enhancement of the decay $K_{\rm short}
\to (\pi\pi)_{I=0}$. The two terms in \eq{m12k} are usually referred to as
\emph{short-distance}\ and \emph{long-distance}\ contributions.  The
long-distance contribution has defied any reliable calculation from first
principles so far. In this humbling situation we can only compare the
experimental value of $\dm_K$ to the short-distance contribution
\beq%
\dm_K^{\rm SD} = \frac{| \bra{K} H^{|\Delta S|=2} \ket{\,\ov{\!K}}|}{m_K} 
  \, .
\label{demk}
\eeq%
In order to compute $\dm_K^{\rm SD}$ we need the hadronic matrix element
\bea%
 \bra{K} Q(\mu_K) \ket{\,\ov{\!K}} & =& 
  \frac{2}{3} M_{K}^2 \, f_K^2 \, 
  \frac{\widehat B_K}{ b_K(\mu_K)} . \label{melk}%
\eea%
Contrary to the situation in the $B$ system, the Kaon decay constant
$f_K=160\,\mev$ is well-measured. We remark here 
that we know $\widehat B_K$ in a particular limit of QCD:
If the number of colours $N_c$ is taken to infinity, $ \bra{K} Q(\mu_K)
\ket{\,\ov{\!K}}$ can be expressed in terms of the current matrix element
$\bra{0}\ov{d}_L \gamma_{\nu} s_L \ket{\,\ov{\!K}}$ which defines $f_K$. For
$N_c=\infty$ one finds $\widehat B_K/ b_K(\mu_K)=3/4$; including certain
calculable (``factorisable'') $1/N_c$ corrections changes this to $\widehat
B_K/ b_K(\mu_K)=1$. A recent lattice calculation finds 
\cite{Antonio:2007pb}%
\bea%
\widehat B_K &=& 0.72 \pm 0.04 \label{bklatt} .
\eea%
The experimental value of the $K_{\rm long}$--$K_{\rm short}$ mass 
difference is \cite{pdg}
\bea%
\dm_K^{\rm exp} &=& (3.483 \pm 0.006)\, \, \mbox{$\mu$eV} 
       \;=\; (5.292 \pm 0.009 ) \cdot 10^{-3} \, \mbox{ps}^{-1} .
\label{dmkexp}
\eea%
Inserting \eqsand{deshs2}{melk} into \eq{demk} gives
\bea%
\frac{\dm_K^{\rm SD}}{\dm_K^{\rm exp}} &=& 
(0.98 \pm 0.22 ) \widehat B_K . \label{sdexp}
\eea%
$\dm_K^{\rm SD}$ is dominated by the term proportional to $ ( V_{cs} V_{cd}^*
  )^2$ and the error in \eq{sdexp} essentially stems from $\eta_{cc}$ in 
\eq{etas}. This uncertainty will shrink when $\eta_{cc}$ is calculated to 
NNLL accuracy. With \eq{bklatt} we find that $ H^{|\Delta S|=2}$ contributes 
$(70\pm 25)$\% to the measured $\dm_K$.   

The off-diagonal element of the decay matrix is given by 
\begin{eqnarray}
\Gamma_{12} & = & \mbox{Abs}\,
    \frac{i}{2 m_K} \int \! d^4 x \,
    \bra{K} H^{|\Delta S|=1} (x)\, H^{|\Delta S|=1} (0) \ket{\,\ov{\!K}}
\label{ga12k} \\
&=& \frac{1}{2 m_K} \sum_f (2\pi)^4 \delta^4 ( p_K-p_f )
  \bra{K} H^{|\Delta S|=1}\ket{f}\, 
  \bra{f} H^{|\Delta S|=1} \ket{\,\ov{\!K}}
  \simeq \frac{1}{2 m_K}\, A_0^*\, \ov{A}_0 \,. \label{g12k}
\end{eqnarray}
Here ``Abs'' denotes the absorptive part of the matrix element. 
$\Gamma_{12}$ is an inclusive quantity built out of all final states $f$ 
into which both $K$ and $\,\ov{\!K}$ can decay. A special feature of the
neutral Kaon system is the saturation of $\Gamma_{12}$ by a single decay
mode, which is $K \to (\pi\pi)_{I=0}$. The notation $A_0$ and $\ov A_0$
for the corresponding decay amplitudes has been introduced after \eq{ek}. 
$\Gamma_{12}$ is a non-perturbative  quantity and its computation on the
lattice involves the difficult task to understand and master 
the $\Delta I=0$ rule. The relation between $\Gamma_{12}$ and $\dg_K$ 
has been derived in \eq{mgsol:b}. Experimentally we have \cite{pdg}
\begin{eqnarray}
\dg_K^{\rm exp} &=&  (7.335 \pm 0.004 ) \, \mbox{$\mu$eV} 
          \; =\; (11.144 \pm 0.006) \cdot 10^{-3} \, \mbox{ps}^{-1} .
\label{dgkexp}
\end{eqnarray}
With \eqsand{dmkexp}{dgkexp} we have precise experimental information 
on $|M_{12}|\simeq\dm_K/2$ and $|\Gamma_{12}|\simeq\dg_K/2$. 
To fully characterise \kkm\ we also need to know the phase $\phi$ defined in 
\eq{defphi}. As in the case of \bbm\ we study a $C\!P$ asymmetry in
a flavour-specific decay mode. With \eqsand{defpq}{defa} one easily finds
\begin{eqnarray}
 A_L &\equiv &
  {\Gamma(K_{\rm long} \to \ell^+\nu\,\pi^-) - 
          \Gamma(K_{\rm long}\to \ell^-\bar\nu\,\pi^+) \over
   \Gamma(K_{\rm long} \to \ell^+\nu\,\pi^-) + 
          \Gamma(K_{\rm long}\to \ell^-\bar\nu\,\pi^+)} \nn
&=& \frac{1 - |q/p|^2}{1 + |q/p|^2}
  \; \simeq \; \frac{a}{2} . \label{al}  
\end{eqnarray} 
At this point it is worthwhile to look back at the quantity $\e_K$ which we
have encountered in the first lecture in \eq{ek}. From \eq{ekqp} we have
learned that $\real \e_K$ measures CP violation in mixing quantified by 
$1-|q/p|$, just as $A_L$ in \eq{al}. While $\imag \e_K$ is related to a
different physical phenomenon, namely mixing-induced $C\!P$ violation, it 
provides the very same information on the fundamental parameters of \kkm:
Since $K \to (\pi\pi)_{I=0}$ dominates $\Gamma_{12}$, 
the $C\!P$-violating phase of $\ov{A}_0/A_0$ equals $\arg \Gamma_{12}$,
see  \eq{g12k}. With this observation and the help of \eq{agmk} we can   
express $\e_K$ in \eq{ekqp} entirely in terms of $\dm_K$, $\dg_K$ and 
$\phi$. Interestingly, the phase $\phi_\e$ of $\e_K$ (see \eq{ek}) is simply 
given by 
\begin{eqnarray}
\phi_{\e} & =  & \arctan \frac{\dm_K}{\dg_K/2} \label{phie} . 
\end{eqnarray}
More details of this calculation can be found in Chapter 1.6 of 
Ref.~\cite{run2}. Nature chose $\dm_K\approx \dg_K/2$ by accident, 
so that $\phi_{\e}$ in \eq{ek} is close to $45^\circ$. The bottom line 
is that $\phi_\e$ carries no information on $C\!P$ violation and 
that $|\e_K|$ and $A_L$ involve the same fundamental $C\!P$-violating 
quantity, which is $\phi$. 
To extract $\phi$ from $A_L$ in \eq{al} or from $\e_K$ in \eq{ekqp} we 
use \eq{agmk}, with
$2|M_{12}|/|\Gamma_{12}|\simeq\dm_K/(\dg_K/2)$ traded for $\tan \phi_\e$:
\begin{eqnarray}
A_L &=& \frac12 \sin (2\phi_\e) \phi + {\cal O} (\phi^2) \nn 
\e_K &\simeq& \frac12  \sin (\phi_\e) e^{i \phi_\e} \phi 
         + {\cal O} (\phi^2) \label{aephi}
\end{eqnarray}
Using the experimental value 
\begin{eqnarray}
A_L^{\rm exp} &=& \lt( 3.32 \pm 0.06 \rt) \times 10^{-3} \no 
\end{eqnarray}
gives 
\begin{eqnarray}
\phi &=& (6.77 \pm 0.12) \times 10^{-3} \label{phikres} .  
\end{eqnarray}
This number is in reasonable agreement with 
$\phi=(6.48 \pm 0.03) \times 10^{-3}$ found from
$\e_K$ with \eq{aephi}.  Next we relate 
$\phi$ to a constraint on $(\ov \rho,\ov \eta)$: Specifying to the 
standard phase convention for the CKM matrix (with $V_{us}V_{ud}^*$ real
and positive) we start from \eq{defphi} to write 
\begin{eqnarray}
   \phi &=& \arg \left( -\frac{M_{12}}{\Gamma_{12}} \right) 
       \; \simeq \; \frac{\imag M_{12}}{|M_{12}|} - 
                    \arg(-\Gamma_{12}) 
       \; =\; 2 \, \lt[ \frac{\imag M_{12}}{\dm_K^{\rm exp}} + \xi_K \rt] 
\label{phimg}
\end{eqnarray}
where 
\begin{eqnarray}
  2 \xi_K &\equiv & - \arg(-\Gamma_{12}) 
          \;\simeq \; 
     - \arg \lt( - \frac{\ov A_0}{A_0} \rt). \label{defxik}
\end{eqnarray}
In \eq{phimg} I have used that the phases of $M_{12}$ and $-\Gamma_{12}$
are separately small in the adopted phase convention and further 
traded $|M_{12}|$ for the experimental $\dm_K/2$. In \eq{defxik} 
the saturation of $\Gamma_{12}$ by $A_0^*\, \ov{A}_0$ in \eq{g12k} 
has been used. Thus $-\xi_K$ is just the CP-odd phase 
in the decay $\ov K \to (\pi\pi)_{I=0}$. 
A recent analysis has estimated $\xi_K \approx -1.7 \cdot 10^{-4}$ 
\cite{bg}, so that $\xi_K$ contributes roughly $-6\%$ to the measured value
of $\phi$. The dominant term  proportional to 
$\imag M_{12}=\imag \bra{K} H^{|\Delta S|=2} \ket{\,\ov{\!K}}$ involves
the CKM factors 
\begin{eqnarray}
  \imag  ( V_{ts} V_{td}^* )^2 &\simeq& 2 (A \lambda^2)^4 \lambda^2
                                     \, \ov\eta \, (1-\ov \rho ) \nn
  \imag  (2 V_{ts} V_{td}^* V_{cs} V_{cd}^*) &\simeq &
  - \imag  ( V_{cs} V_{cd}^* )^2  \simeq 2 (A \lambda^2)^2 \lambda^2
  \, \ov \eta,  \label{theimags}
 \end{eqnarray}
 where the lowest-order Wolfenstein expansion has been used. Inspecting
 the dependences of the CKM factors on $\ov\rho$ and $\ov\eta$ we see
 that the experimental constraint from $\phi$ defines a hyperbola in the
 $(\ov \rho,\ov \eta)$ plane.  Combining \eq{theimags} with
 \eqsand{deshs2}{phimg}, inserting the QCD factors from \eq{etas} and
the matrix element from \eq{melk} and finally 
 using $\phi=(6.48 \pm 0.03)\times 10^{-3}$ from $\e_K$ this hyperbola reads
\begin{eqnarray}
\ov \eta &=& \frac{1}{\widehat B_K}  \, 
 \frac{0.34\pm 0.03}{1.3\pm 0.1 -\ov \rho} 
\label{hypres} .
\end{eqnarray}
The uncertainties in $\widehat B_K$ from \eq{bklatt} and from 
$\eta_{cc}$ and $\eta_{ct}$ in \eq{etas} (reflected by $1.3\pm 0.1$) 
inflict errors of similar size on the $\ov \eta$ extracted from 
\eq{hypres}. The numerator $0.34\pm 0.03$ is calculated with
$|V_{cb}|=A\lambda^2 = 0.0412\pm 0.0011 $. The 10\% uncertainty 
of this number stems solely from the error in $|V_{cb}|$, which enters 
$\ov\eta$ in \eq{hypres} with the fourth power.

The neutral Kaon system is the only neutral meson
system for which all three quantities $\dm$, $\dg$ and $\phi$ are measured.
It should be stressed that also the sign of $\dg/\dm$ is firmly established.
Measuring $\mbox{sign}\,(\dg/\dm) $ is difficult for all \mm\ systems.  In the
neutral Kaon system the measurement of $\dm$ and  $\mbox{sign}\,(\dg/\dm) $
 uses \emph{$K_{\rm short}$
  regeneration}: If a  $K_{\rm long}$ beam hits a nucleus in a target 
(the regenerator), strong inelastic scattering changes the 
$\ket{K_{\rm long}}$ state into a superposition of $\ket{K_{\rm long}}$
and  $\ket{K_{\rm short}}$ giving access to observables which are sensitive 
to $\dm$ and the abovementioned sign. For details on these experimental 
aspects I refer to \cite{mrr}.
 
Finally I discuss \ddm: Box diagrams in \fig{fig:boxes} with one or two
internal $b$ quarks are highly CKM-suppressed. The dominant box diagrams with
internal $d$ and $s$ quarks suffer from a very efficient GIM suppression
proportional to $m_s^4/m_c^2$. This makes the diagrams sensitive to very low
scales and perturbative calculations of $\dm_D$, $\dg_D$ and $a_{\rm fs}^D$
are put into doubt. In the effective theory both $M_{12}$ and $\Gamma_{12}$
are dominated by the bilocal contribution with $H^{|\Delta C|=1}$.  The only
possible clear prediction is the qualitative statement that all these
quantities are very small.  Theoretical calculations usually quote numbers for
the quantities $x\equiv \dm_D/\Gamma_D$ and $y\equiv \dg_D/(2 \Gamma_D)$. The
theoretical predictions for $|x|,|y|$ cover the range from zero to
$|x|,|y|\sim 0.01$ and come without reliable error estimates. Therefore
current experimental values are compatible with the SM but may also be
dominated by a new physics contribution. A ``smoking gun'' of new physics,
however, would be the discovery of a non-zero CP asymmetry in the $D$ system.
  
\boldmath
\subsection{Mixing-induced $C\!P$ asymmetries}
\unboldmath
\parbox[b]{0.55\textwidth}{
At the end of Sect.~\ref{sec:time} we have learned that mixing-induced $C\!P$
asymmetries can provide clean information on fundamental $C\!P$ phases in the
Lagrangian. These  $C\!P$ asymmetries involve the interference between mixing
and decay amplitudes as depicted on the right.} \hfill
\parbox[b]{0.3\textwidth}{
 ${B}\qquad \stackrel{\ds q/p}{\mbox{\LARGE $\longrightarrow$}}
         \qquad {\Bbar}$\\[2mm]
$A_f\!\!$
     {\Large $\searrow$} \hspace{1cm} {\Large $\swarrow$}$\,\ov{\!A}_f$
     \\[2mm]
\mbox~\hspace{1.5cm} $f$\\[-2mm]}~~~~~~~~\\[3mm]
In this lecture we restrict the discussion to gold-plated modes which involve 
a $C\!P$ eigenstate $f_{\rm CP}$ in the final state, cf.\
\eq{defcpeig}.\footnote{One
can also identify gold-plated decays into non-CP eigenstates, important
channels are e.g.\ $B_s\to D_s^\pm K^\mp $.} 
In the $B_d$ and $B_s$ meson systems the  mixing-induced $C\!P$ asymmetries are
a real gold mine, because there are many decay modes satisfying the condition
for a golden decay mode as defined after \eq{dirmix}. Prominent examples are
the decays $B_s\to J/\psi \phi$ and $B_d\to J/\psi K_{\rm short}$, whose decay
amplitudes essentially only involve the CKM factor $V_{cs} V_{cb}^*$. To
understand this first note that the decay proceeds at tree--level by
exchanging a W boson. There are also contributions involving an up, charm or
top quark loop, with attached gluons splitting into the charm-anticharm pair
hadronising into the $J/\psi$ meson. Such diagrams are called \emph{penguin
  diagrams}. A penguin diagram in the narrow sense only involves one neutral
vector boson (which can be a gluon, photon or Z boson). A gluonic penguin
diagram is depicted in \fig{fig:peng}. (Yet a $J/\psi$ cannot be
produced from a single gluon. One needs a photon or three gluons at least.)
\begin{nfigure}{t}
\centering 
\includegraphics*[scale=0.7,angle=-90]{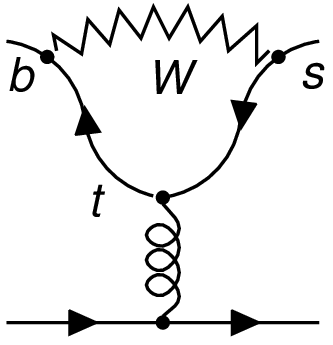}
\caption{Gluonic penguin diagram with an internal top quark.}\label{fig:peng}
\end{nfigure}
In the context of  mixing-induced $C\!P$ asymmetries one often speaks of 
\emph{penguin pollution}, because the penguin diagrams may involve different 
CKM factors than the tree diagram spoiling the golden-mode property. To
estimate the  penguin pollution in  $B_s\to J/\psi \phi$ and $B_d\to J/\psi
K_{\rm short}$ first use the unitarity relation $V_{ts} V_{tb}^*=
-V_{cs} V_{cb}^* - V_{us} V_{ub}^*$ to write 
\begin{eqnarray}
H^{|\Delta B|=1} &=& \phantom{+} 
              V_{cs} V_{cb}^* h_c + V_{cs}^* V_{cb} h_c^\dagger 
   \, +\, V_{us} V_{ub}^* h_u 
             +         V_{us}^* V_{ub}  h_u^\dagger 
       . \label{hcu} 
\end{eqnarray}
Here the last two terms are highly suppressed, since $|V_{us}^* V_{ub}|
\sim 0.03 \, |V_{cs}^* V_{cb}|$. Moreover, $h_u$ has no tree
contributions, but solely stems from penguin diagrams with up and top
quarks. Since these loop effects involve non-perturbative physics, it is
difficult to quantify the loop suppression. Still, the CKM suppression
is efficient enough to render the modes gold-plated at the level of a
few percent. Since the CKM elements are factored out in \eq{hcu}, $h_c$
and $h_u$ only contain Wilson coefficients, operators and real
constants. Importantly, $h_{u,c}$ and $h_{u,c}^\dagger$ are related by
the $C\!P$ transformation:
\begin{eqnarray}
     h_{u,c}^\dagger &=&  (CP)^\dagger h_{u,c} C P  .\label{cph}
\end{eqnarray}
While I discuss $B_s\to J/\psi \phi$ and $B_d\to J/\psi K_{\rm short}$ here
for definiteness, the results apply to other gold-plated $M\to f_{\rm CP}$
modes as well, with obvious replacements for the CKM elements.
The underlying reason for the cancellation of hadronic uncertainties in
gold-plated decays is the $C\!P$ invariance of QCD: While we cannot 
compute $\bra{f_{\rm CP}} h_c \ket{B} $\footnote{In flavour physics
  matrix elements like $\bra{f} H^{|\Delta B|=1} \ket{M}$ are always
  understood to include the strong interaction. This means that the
  fields are understood as interacting fields in the Heisenberg picture
  with respect to the strong interaction.}, we can relate this matrix element
to $\bra{f_{\rm CP}} h_c^\dagger \ket{\,\ov{\!B}} $ through
\begin{eqnarray}
     \bra{f_{\rm CP}} h_{u,c}^\dagger \ket{\,\ov{\!B}} &=& 
     \bra{f_{\rm CP}} (CP)^\dagger h_{u,c} CP
                \ket{\,\ov{\!B}}
 \;=\; - \eta_{\rm CP} 
     \bra{f_{\rm CP}} h_{u,c} \ket{B}
, \label{cpdec}
\end{eqnarray}
where I just used the $C\!P$ transformations of \eqsto{defcpm}{defcpeig} and
\eq{cph}. We first apply this to the decay mode $B_s\to J/\psi \phi$.  The
final state consists of two vector mesons. By conservation of angular momentum
they can be in states with orbital angular momentum quantum numbers $l=0,1$ or
$2$: The two spin-1 states of the vector mesons can be added to a state of
total spin $0,1$ or 2, which requires an orbital angular momentum 
of $l=0,1$ or $2$ to give a $
J/\psi \phi$ state with zero total angular momentum.  The p-wave state with
$l=1$ is $C\!P$-odd and the other two states are $C\!P$-even, owing to the
parity quantum number $(-1)^l$ of their spatial wave function. Experimentally
one separates these states by an angular analysis \cite{ddlr,dfn} of the data
sample. This can be done including the full time dependence of the decay, so
that we can isolate the time-dependent $C\!P$ asymmetries in the different
partial-wave channels.  The most-populated state is the CP-even $l=0$ (i.e.\ 
s-wave) state.  Writing $f_{\rm CP}=(J/\psi \phi)_l$ with $\eta_{\rm
  CP}=(-1)^l$ we obtain for the amplitudes $A_{f_{\rm CP}}$ and $\ov A_{f_{\rm
    CP}}$ (see \eq{defaf}):
\begin{eqnarray}
    \frac{\ov A_{f_{\rm CP}}}{A_{f_{\rm CP}}} &\simeq& 
      \frac{ \bra{f_{\rm CP}} H^{|\Delta B|=1}  \ket{\,\ov{\!B}_s}}{
             \bra{f_{\rm CP}} H^{|\Delta B|=1}  \ket{B_s}} 
    \; =\; \frac{V_{cs}^* V_{cb}}{V_{cs} V_{cb}^*} 
      \frac{ \bra{f_{\rm CP}} h_{c}^\dagger \ket{\,\ov{\!B}_s}}{
             \bra{f_{\rm CP}} h_{c} \ket{B_s}}  
    \; =\; - \eta_{\rm CP} \, \frac{V_{cs}^* V_{cb}}{V_{cs} V_{cb}^*} 
\end{eqnarray}
Combining this result with \eqsand{qpb}{deflaf} we find 
\begin{eqnarray}
       \lambda_{f_{\rm CP}} &=& \eta_{\rm CP} \,
      \frac{V_{tb}^* V_{ts}}{V_{tb} V_{ts}^*}
      \frac{V_{cs}^* V_{cb}}{V_{cs} V_{cb}^*} \; = \; 
   \eta_{\rm CP} \, e^{2 i \beta_s}. \label{atfcp}
\end{eqnarray}
In the last step I have used the definition of $\beta_s$ in \eq{betasdef}. 
\begin{nfigure}{t}
\centering 
\includegraphics*[scale=0.7,viewport=100 540 530 700]{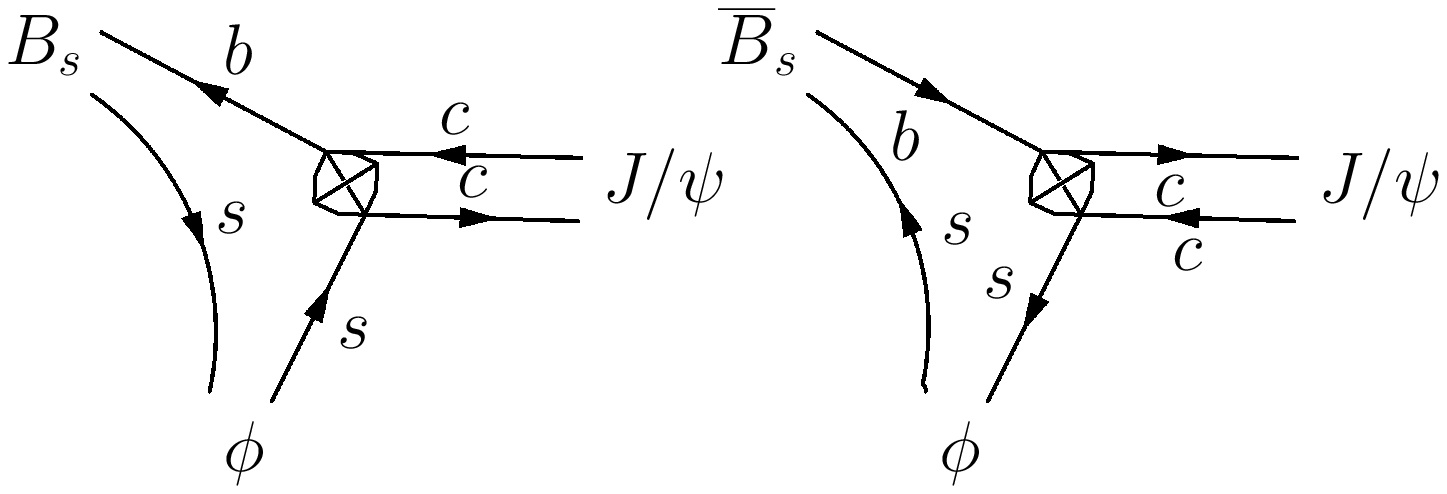}\\
\includegraphics*[scale=0.7,viewport=100 540 530 700]{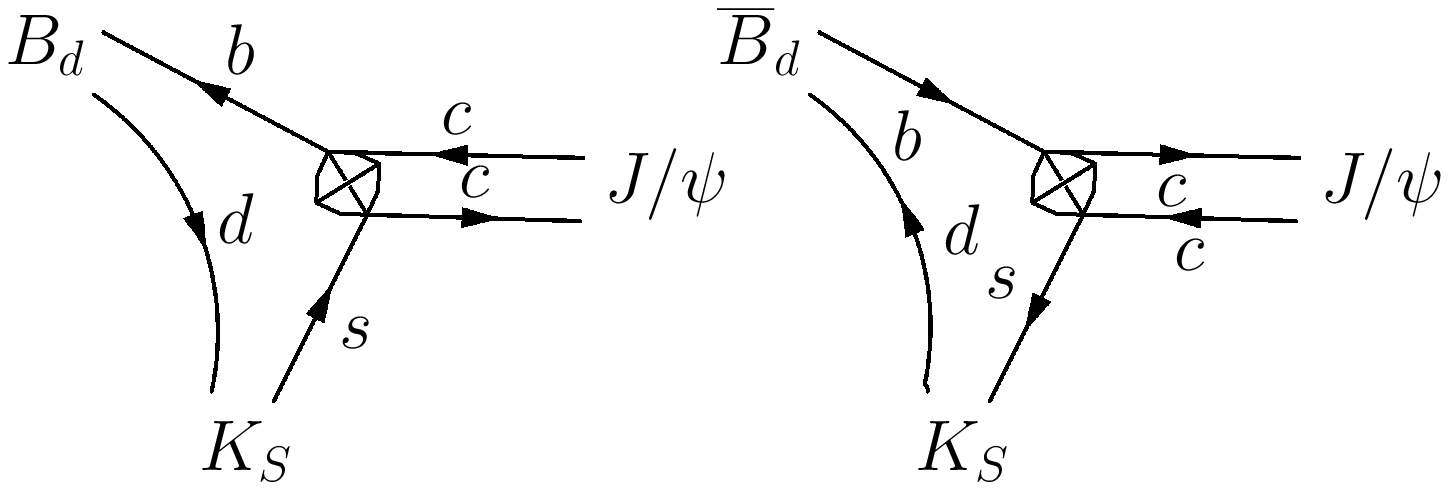}
\caption{Interfering amplitudes which give rise to mixing-induced CP 
violation for the two golden modes discussed in the
  text.}\label{fig:gold}
\end{nfigure}
With \eq{atfcp} we can calculate the time-dependent $C\!P$ asymmetry of
\eq{acp}. First we verify that our golden mode satisfies $|\lambda_{f_{\rm
    CP}}|=1$, so that $A_{CP}^{\rm dir}$ in \eq{dirmix} vanishes. 
The other two quantities in \eq{acp} evaluate with \eq{atfcp} to 
$A_{CP}^{\rm mix}= -\eta_{\rm CP} \sin (2 \beta_s)$ and 
$A_{\dg} = -\eta_{\rm CP} \cos(2 \beta_s)$, so that  
(neglecting the tiny ${\cal O}(a) $ term)
\begin{eqnarray}
a_{f_{\rm CP}}(t) = \eta_{\rm CP} \frac{\sin (2 \beta_s) 
       \sin ( \dm_{B_s} \, t )}{
       \cosh (\dg_{B_s} \, t/ 2) -
       \eta_{\rm CP} \cos(2 \beta_s) \sinh  (\dg_{B_s} \, t / 2) }
\qquad\mbox{for }f_{\rm CP}=(J/\psi\phi)_l 
. \label{acpbs}%
\end{eqnarray}
In the SM $\beta_s$ is small and $a_{(J/\psi\phi)_l}(t)$ is an ideal 
testing ground to find new physics \cite{ln,dfn}. 

Next I discuss $B_d \to J/\psi K_{\rm short}$. The final state has
orbital angular momentum $l=1$ balancing the spin of the $ J/\psi$.
Neglecting the small $C\!P$ violation in \kkm\ we can regard the $K_{\rm
  short}$ as $C\!P$-even. The $J/\psi$ is $C\!P$-even as well and the
orbital angular momentum contributes a factor of $-1$ to the total
$C\!P$ quantum number. Thus $\eta_{J/\psi K_{\rm
    short}}=-1$.  From \fig{fig:gold} we observe a novel feature
compared to $B_s\to J/\psi \phi$. The interference of the $B_d$ and
$\Bbar_d$ decays involves \kkm: The $B_d$ decay involves the $K$
component of $K_{\rm short}$, while the $\Bbar_d$ decays into the
$\Kbar$ component of $K_{\rm short}$.  Experimentally the $K_{\rm
  short}$ is detected via a  pair of charged pions 
whose invariant mass equals $M_K$, denoted here by $(\pi^+\pi^-)_K$. 
Therefore we should identify the amplitudes $A_{f_{\rm
    CP}=J/\psi K_{\rm short}}$ and $\ov A_{f_{\rm CP}=J/\psi K_{\rm
    short}}$ with $ A(B_d \to J/\psi K \to J/\psi (\pi^+\pi^-)_K)$ and $
\ov A(\Bbar_d \to J/\psi \ov K \to J/\psi (\pi^+\pi^-)_K)$,
respectively.  Therefore
\beq%
\frac{\ov A_{J/\psi K_{\rm short}}}{A_{J/\psi K_{\rm short}}} =
\frac{V_{cb}V_{cs}^*}{V_{cb}^*V_{cs}} \frac{V_{us}V_{ud}^*}{V_{us}^*V_{ud}},
\qquad\qquad \lambda_{J/\psi K_{\rm short}} = - \frac{V_{tb}^{*}V_{td}}{V_{tb}
  V_{td}^{*}} \frac{V_{cb}V_{cs}^*}{V_{cb}^*V_{cs}}
\frac{V_{us}V_{ud}^*}{V_{us}^*V_{ud}} \simeq - e^{-2 i \beta}
\eeq%
In the last step I have used the definition of $\beta$ in \eq{eq:beta} and
neglected $\arg[-V_{cd}V_{cs}^*/(V_{ud}V_{us}^*)]\simeq A^2 \lambda^4 \ov \eta
< 10^{-3}$, so that $\imag \lambda_{J/\psi K_{\rm short}}\simeq
\sin(2\beta)$. We may further neglect $\dg_{B_d}$ in \eq{acp} to find
the most famous time-dependent $C\!P$ asymmetry,
\begin{equation}
a_{J/\psi K_{\rm short}}(t) = \sin(2\beta) \sin(\dm_{B_d} t) . 
\label{famous}
\end{equation}

Finally I give a (very incomplete) list of other golden $M\to f_{\rm
  CP}$ decays. The decay $B_s \to J/\psi \phi$ can be substituted for
$B_s \to J/\psi \eta^{(\prime)}$, which does not require any angular
decomposition. In a hadron collider experiment $\eta$'s and
$\eta^\prime$'s are hard to detect, but $B_s \to J/\psi \eta^{(\prime)}$
is interesting for B factories running on the $\Upsilon(5S)$ resonance.
While the modes discussed above provide insight into the physics of
\bbm, one can also use mixing-induced $C\!P$ violation to probe $C\!P$
phases from new physics in loop-induced $B$ decays such as $B_d\to \phi
K_{\rm short}$ \cite{gw}. This mode is triggered by the quark decay $\ov
b\to \ov s s \ov s $. The same transition in probed in $B_s \to \phi
\phi$. Likewise new physics in the $\ov b \to \ov s d \ov d$ amplitude
may reveal itself in $B_s \to K_{\rm short} K_{\rm short}$. 
Gold-plated $D^0$ decays are $D^0 \to K_{\rm short} \pi^0$ and $D^0 \to K_{\rm
  short} \rho^0$, which are penguin-free $c\to s \ov d u $ decays. A
gold-plated $K$ decay is $K_{\rm long}\to \pi^0 \nu \ov \nu $ \cite{bb}. Here
no \mm\ oscillations are present, but \kkm\ nevertheless enters the process
through the mass eigenstate $K_{\rm long}$.  The final state $ \pi^0 \nu \ov
\nu$ is $C\!P$-even and the dominant contribution to the decay involves
mixing-induced $C\!P$ violation, i.e.\ the decay amplitude is proportional to
$\imag\lambda_f$ (see e.g.\ Ref~\cite{nir}).

\subsection{The unitarity triangle}
Many measurements contribute to the global fit of the unitarity triangle
defined in \eq{defre} and depicted in \fig{fig:ut}. Conceptually it is
useful to disentangle tree decays from FCNC processes: 
Tree-level amplitudes are insensitive
to new physics and therefore determine the true apex $(\ov \rho,\ov
\eta)$ of the unitarity triangle. In principle one could determine the
unitarity triangle in this way, insert the result into the SM
predictions of the FCNC processes and then assess the possible impact of
new physics on the latter. In practice, however, the tree constraints
still suffer from large uncertainties, while for example $a_{J/\psi
  K_{\rm short}}(t)$ in \eq{famous} and $\dm_{B_d}/\dm_{B_s}$ in
\eq{home} determine $\sin(2 \beta)$ and the side $R_t$ 
(see \eq{defsd}) fairly precisely. 
Therefore, for the time being, it is best to combine all information 
into a global fit of the unitarity triangle. 

From $b\to c \ell \ov\nu$ decays $|V_{cb}|\simeq A\lambda^2$ is
precisely determined. Therefore we realise from \eqsand{defsd}{vub} that
any measurement of $|V_{ub}|$ essentially fixes the side $R_u$ of the
triangle. $|V_{ub}|$ is determined from (inclusive or exclusive)
semileptonic $b\to u$ decays and hadronic uncertainties limit the
accuracy of the extracted $|V_{ub}|$ to 8-10\%. The theoretical
methods used to determine $|V_{cb}|$ and   $|V_{ub}|$
are briefly reviewed in Ref.~\cite{lp05}. The angle $\gamma$ of the
unitarity triangle is currently measured in two ways from tree-level
decays: First, the interference of the 
$b\to c \ov u s$ and $b\to u \ov c s$ amplitudes in 
$B^\pm \to \DorDbar\, K^\pm$ decays is exploited \cite{glw}.  
Second, one measures mixing-induced $C\!P$ violation in 
$B_d \to \pi \pi$, $B_d \to \rho \pi$ or $B_d \to \rho \rho$ decays, 
which allows to find the angle $\alpha$ of the desired triangle.  
These modes are not gold-plated and suffer from penguin pollution, 
which, however, can be eliminated by means of an isospin analysis 
\cite{glo}. While the extracted result for $\alpha$  is sensitive to new
physics in \bbmd, this possible effect can be eliminated if the 
measured $\alpha^{\rm exp}$ and $\beta^{\rm exp}$ are combined to 
give $\gamma^{\rm exp} = \pi -\alpha^{\rm exp} -\beta^{\rm exp}$. 
Combining the constraints from $|V_{ub}|$, $\gamma$ and $\alpha$ with
those from \mmm\ discussed in this lecture results in the unitarity
triangle shown in \fig{fig:utfit}. 
\begin{nfigure}{t}
\centering 
\includegraphics*[scale=0.65,angle=0]{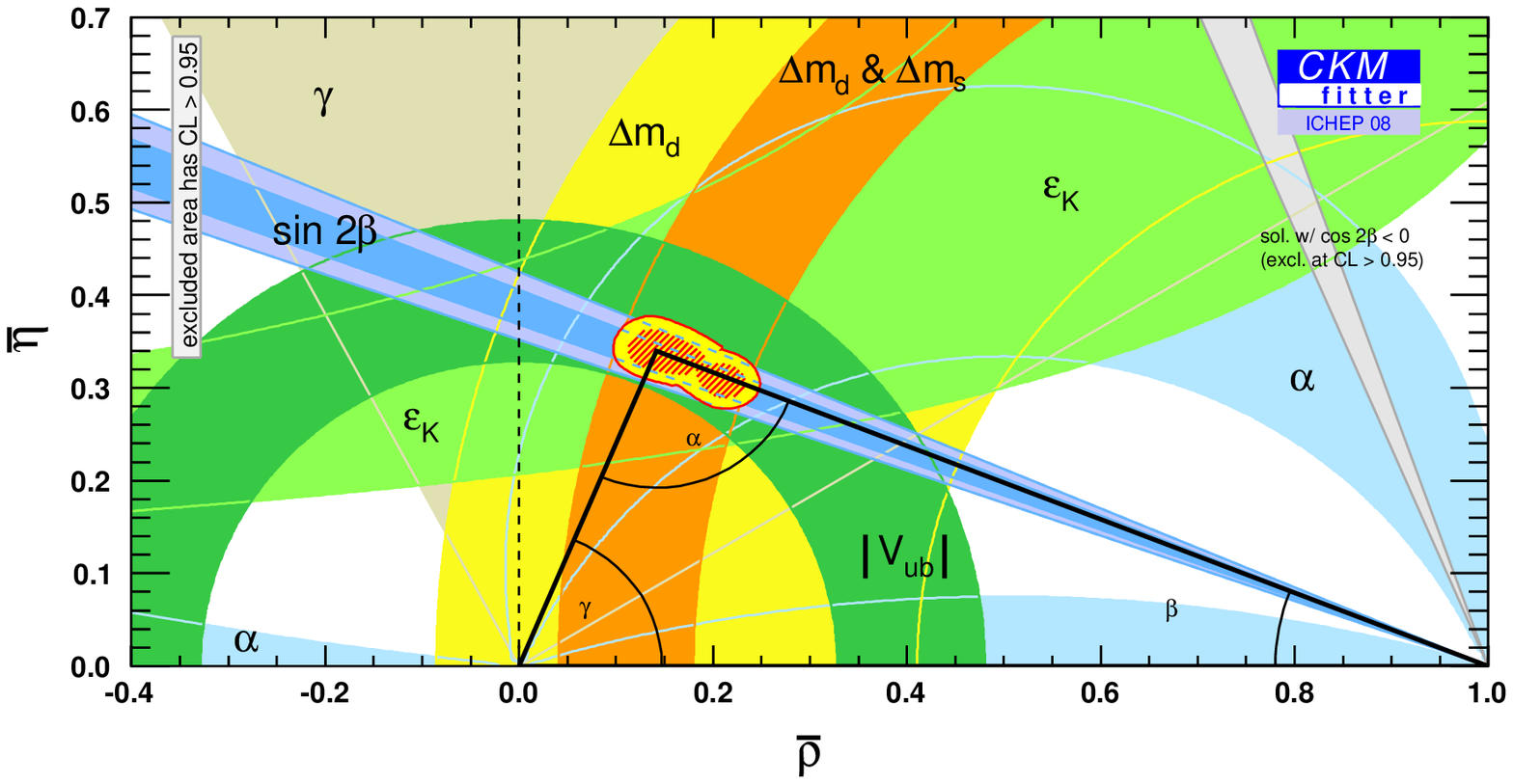}
\caption{Global fit to the unitarity triangle from the CKMFitter group 
\cite{ckmf}. A different statistical approach is used by the UTFit group
\cite{utf}.}\label{fig:utfit}
\end{nfigure}

\section*{Suggestions for further reading}
There are many good review articles on \mmm\ and flavour physics in
general, putting emphasis on different aspects of the field.  A student
interested in the theoretical foundation of flavour physics, effective
Hamiltonians and higher-order calculations is referred to the lecture in
Ref.~\cite{Buras:1998raa} and the review articles in
Refs.~\cite{run2,bbl}. Most reviews and lectures focus on CP violation
and I recommend Refs.~\cite{Nir:2005js} and \cite{Fleischer:2004xw}.  I
have only briefly touched \ddm, two review articles dedicated to $D$
physics are cited in Ref.~\cite{charmrev}. Lectures covering both $K$
and $D$ physics can be found in Ref.~\cite{Buchalla:2001ux}.  A concise
summary of the physics entering CKM metrology can be found in
Ref.~\cite{lp05}, a more elaborate article on the subject is
Ref.~\cite{Ali:2003te}.  Standard textbooks on flavour physics are
listed in Ref.~\cite{books}.

\section*{Acknowledgements} 
I am grateful to Ahmed Ali and Misha Ivanov for the invitation to this summer
school. It has been a pleasure to discuss so many different fields of physics
with the other lecturers and the participating students.  I thank Momchil
Davidkov, Lars Hofer and Dominik Scherer for proofreading this text.

\end{document}